\documentclass{aa}
\usepackage{graphicx}
\usepackage{txfonts}

\begin{document} 

   \title{A young spectroscopic binary in a quintuple system part of the Local Association.}


   \author{
                  Carlos~Cardona~Guill\'en\inst{1, 2}
          \and
          Nicolas~Lodieu\inst{1, 2}
          \and
          V\'ictor~J.~S.~B\'ejar\inst{1, 2}
          \and
          David~Baroch\inst{3, 4}
          \and
          David~Montes\inst{5}
          \and
          Matthew~J.~Hoskin\inst{6, 7}
          \and
          Sandra~V.~Jeffers\inst{8}
          \and
          Felipe Murgas\inst{1,2}
          \and
          Pier-Emmanuel~Tremblay\inst{6}
          \and
          Patrick~Sch\"ofer\inst{8}
          \and
          Daniel~Harbeck\inst{9}
          \and
          Curtis~McCully\inst{9}
          }

   \institute{
                Instituto de Astrof\'isica de Canarias, V\'ia L\'actea s/n, E-38205 La Laguna, Tenerife, Spain
                \and
                Departamento de Astrof\'isica, Universidad de La Laguna, E-38200, La Laguna, Tenerife, Spain
                \and
                Institut de Ci\`encies de l'Espai (ICE, CSIC), Campus UAB, c/Can Magrans s/n, E-08193 Bellaterra, Barcelona, Spain
                \and
                Institut d'Estudis Espacials de Catalunya (IEEC), c/ Gran Capit\`a 2-4, E-08034 Barcelona, Spain
                \and
                Departamento de F{\'i}sica de la Tierra y Astrof{\'i}sica \& IPARCOS-UCM (Instituto de F\'{i}sica de Part\'{i}culas y del Cosmos de la UCM), Facultad de Ciencias F{\'i}sicas, Universidad Complutense de Madrid, E-28040 Madrid, Spain
                \and
                Department of Physics, University of Warwick, Coventry, CV4 7AL, UK
                \and
                Centre for Exoplanets and Habitability, University of Warwick, Coventry, CV4 7AL, UK
                \and
                Max-Planck-Institut f\"ur Sonnensystemforschung, Justus-von-LiebigWeg 3, 37077 G\"ottingen, Germany
                \and
                Las Cumbres Observatory Global Telescope Network, 6740 Cortona Dr. Suite 102, Goleta, CA 93117
        }

   \date{\today{}}

 
  \abstract
   {
    Double-lined spectroscopic (SB2) binaries allow us to obtain a direct determination of the masses of their components, which is essential to test stellar models.
    Although these objects only provide a lower limit for the mass, they are more abundant than their eclipsing counterparts as they are not as strongly limited by the inclination of their orbit.
   }
   {
        Our aim is to derive the orbital and physical parameters of GJ\,1284, a young SB2.
        We also revise the membership of this system and its two wide co-moving companions, GJ\,898 and GJ\,897AB, to a young moving group to assess, along with other youth indicators, their age.
        Afterwards, we compare the results from these analyses and the photometry of these systems with several pre-main-sequence evolutionary models.
   }
   {
        We use high-resolution spectra to determine the radial velocity of each component of GJ\,1284 and the orbit of the system alongside its systemic velocity.
        Additionally, we use TESS photometry to derive the rotational period of the GJ\,1284 and its two wide companions.
   }
   {
                GJ\,1284 is a binary system located at approximately 16\,pc with an eccentric orbit ($ e = 0.505 $)  of 11.83\,d period made up of  an M2--M2.5\,+\,M3--M3.5 with minimum masses of  $ M\sin ^3 i =  0.141 \pm 0.003 $ and $ 0.1189 \pm 0.003\,\mathrm{M_{\sun}}$, respectively.
        The revised systemic velocity of $ \gamma = 0.84 \pm 0.14\,\mathrm{km\,s}^{-1} $ suggests that it is a member of the Local Association.
        The kinematics together with other activity and youth indicators imply an age of 110--800\,Myr for this system and its two companions.
   }
   {
        The isochronal ages derived from the comparison of the photometry with several evolutionary models are younger than the age estimated from the activity indicators for the three co-moving systems.
        The masses for the components of GJ\,1284, derived from their luminosity and age using the different models, are not consistent with the masses derived from the photometry, except for the PARSEC models, but are compatible with dynamical masses of double-lined eclipsing binaries with similar ages and spectral types.
        The effect of magnetic activity in the form of spots can reconcile to some extent the photometric and dynamical masses, but is not considered in most of the evolutionary models.
   }


   \maketitle


\section{Introduction}
The observational knowledge of fundamental properties of the stars is the basis for most of the astrophysics models explaining the evolution of stars and galaxies.
In particular, the mass of the star is the main property that regulates its evolution and eventual fate.
The determination of stellar masses with high precision is then essential to constrain these models.
Double-lined eclipsing binaries (EBs) provide a unique opportunity to obtain direct measurements of the mass and radius of its stellar components with accuracies of $ \sim 1\% $ \citep{Torres2010}.
However, the number of  systems where this precision has been achieved is relatively low \citep{Torres2010,Southworth2015}.
This holds true,  especially, for young ($ <800 $\,Myr) low-mass stars and brown dwarfs, as only a few are known and have accurate mass and radius determinations \citep[and references therein]{David2019,Lodieu2020a}.
These systems are necessary to constrain pre-main-sequence (PMS) evolutionary models.
Most of them have been recently discovered, made possible by  the advent of ground-based surveys such as SuperWASP \citep{Pollacco2006}, and space missions such as CoRoT \citep{Baglin2006}, Kepler \citep{Koch2004,Borucki2007}, or  TESS \citep{Ricker2014}, which monitor almost the whole night sky.

The detection of EBs is heavily constrained by the inclination of the orbit of the system (i.e. only nearly edge-on configurations produce eclipses).
On the other hand, orbital configurations of double-lined spectroscopic binaries (SB2) are not as strongly limited.
Thus, the number of known young SB2s \citep[e.g.][and references therein]{Dudorov2016,Shkolnik2017,Zuniga-Fernandez2021} is higher than EBs. 
Most of these young SB2s were discovered, also serendipitously, as part of spectroscopic surveys focusing on   relatively few members of young stellar associations,   the majority of them   solar-type (FGK) stars, while the number of M dwarfs still remains low.
Moreover, their orbits, for the most part, have not been characterised to date as only a few spectra for each of them are available.
However, SB2s do not provide any information about the radii of the system components and only a minimum limit for their masses, even though they are  more abundant than EBs.

The determination of the age of the system is a step needed to test the validity of the predictions of PMS models.
This is usually achieved comparing the photometry of the object against  isochrones derived from evolutionary models \citep[e.g.][]{Malo2014b,Herczeg2015,Bell2015}.
In open clusters and young moving groups (YMGs), we can assume that all the members were born roughly at the same time \citep{Zuckerman2004}.
Under this assumption we can expect that the age derived from the isochrones should be consistent for all of them.
This provides a powerful tool for determining the age of a particular open cluster or YMG, and subsequently of any of its newly identified members.
For this reason, a proper characterisation of the evolutionary models is also important for determining stellar ages.
Nonetheless, the age of the members of open clusters and YMGs can also be estimated using other methods that are mostly independent or do not rely so heavily on evolutionary models, for example
traceback simulations to find the convergence point;
lithium depletion boundary (LDB) determinations;
or age relations with youth indicators such as rotational period, X-ray, or H$ \alpha $ emission \citep[see][and references therein]{Soderblom2010}

The relative proximity to the Sun of most YMG members means they are bright enough to achieve a radial velocity (RV) precision of $\mathrm{km\,s^{-1}}$   with small telescopes ($\sim$1\,m)  needed to measure dynamical masses.
As part of an observing campaign with the Mercator 1.2\,m telescope, we observed several early- to mid-type M dwarfs previously classified as members of YMGs in the literature.
We confirmed one of these targets, GJ\,1284, as an SB2 previously identified by \cite{Gizis2002}.
Later on, we also characterised the orbit of this system using the Mercator spectra and data from other facilities obtained during this and subsequent campaigns, and data gathered  from the literature.
The system was initially classified as a member of either the Beta Pictoris Moving Group \citep[$\beta$PMG;][]{BarradoyNavascues1999b,Zuckerman2001} or the Columba Association \citep[COL;][]{Torres2008} by \citet{Malo2013}.
Moreover,  one other system, GJ\,898, had been identified kinematically as a wide co-moving companion to GJ\,1284 \citep{Shaya2011}. 
This wide companion forms part of a triple system alongside another binary star, GJ\,897AB  \citep{Kuiper1943}.
The three objects combined form a wide quintuple system.
In this work we revise the membership of GJ\,1284 to a YMG based on its new kinematics and youth indicators, which suggests that it belongs to the Local Association \citep[LA;][]{Eggen1975,Montes2001}, the same YMG as GJ\,898 and GJ\,897 \citep{Montes2001,Makarov2008}.

The LA, also commonly known as the Pleiades moving group, is one of the first classical YMGs identified by \citet{Eggen1975} who initially described it as a stream associated with the Pleiades open cluster.
Later on, \citet{Asiain1999} and \citet{LopezSantiago2006} revealed that LA is actually made up of several subgroups with ages ranging from 10--300 Myr.

In this work we  characterise the orbit GJ\,1284 using spectra collected with four different high-resolution spectrographs, and compare the results against the predictions made by several stellar evolution models.
In Sect.~\ref{DescriptionSection} we introduce the GJ\,1284 system and review the literature about this object and its wide companions.
In Sect.~\ref{ObservationsSection} we describe the high-resolution spectra collected over several observing campaigns complemented by additional photometric and spectroscopic datasets from the literature.
In Sect.~\ref{DataAnalysisSection} we measure the RVs from the spectra and determine the orbital parameters of the system.
We also infer the rotational period of GJ\,1284 system and its wide companions.
Finally, in Sect.~\ref{DiscussionSection} we propose an age range for  the three systems and compare their photometry and the mass of the GJ\,1284 components against PMS evolutionary models  and similar studies.


\section{GJ\,1284} \label{DescriptionSection}
\subsection{Description of the system}
GJ\,1284 is located at a distance $d=15.906\pm0.019$\,pc \citep[{\it Gaia} early Data Release 3, eDR3;][]{Prusti2016,Brown2020}.
It has been extensively analysed as a member of the solar neighbourhood due to its proximity, and has been part of several direct-image searches for substellar companions \citep[e.g.][]{Galicher2016,Meshkat2017,Naud2017,Nielsen2019}.
The system was classified as an M3.0V star by \citet{Hawley1996}.
More recently, \citet{Torres2006} classified it as an M2.0Ve with an uncertainty of one subclass using high-resolution spectra.
In this work we estimate the spectral types of the binary components based on the colours derived from the photometric magnitudes. 
        
\citet{Gizis2002} measured short-term variations in the RVs of GJ\,1284 of over $ 46\,\mathrm{km\,s^{-1}} $ and identified it as an SB2, even though they did not characterise its orbit.
Later it was again identified as an SB2 by \citet{Torres2006}, who obtained three more RV measurements with a mean value of $ -5.7\,\mathrm{km\,s^{-1}}$.
\citet{Jeffers2018} also measured the RV of GJ\,1284 on seven consecutive nights obtaining  a mean value of $-4.4\,\mathrm{km\,s^{-1}}$  with a peak-to-peak variation  of $62.63\,\mathrm{km\,s^{-1}}$.
Additionally, \citet{Schneider2019} measured a single value of $ 8.18\pm1.07\,\mathrm{km\,s^{-1}}$.
Finally, \citet{Sperauskas2019} obtained two more values of $ -0.4 \pm 1.0 $ and $ -14.6 \pm 1.0\,\mathrm{km\,s^{-1}} $ .
A summary of all the individual measurements is given in Table~\ref{literatureRVs}.
Nonetheless, the orbital parameters of this binary system have not been derived due to the lack of continuous monitoring.

\begin{table}
        \caption{Compilation of RVs measured for GJ\,1284 from the literature.}
        \label{literatureRVs}
        \centering
        \begin{tabular}{c l}
                \hline\hline
                \noalign{\smallskip}
                        RV & Ref  \\\relax
                        [km\,s$ ^{-1} $] &  \\
                \noalign{\smallskip}
                \hline
                \noalign{\smallskip}
                        $ +11.7 \pm 1.5 $\tablefootmark{a} & Giz2002 \\
                        $ -4.7 \pm 1.5  $\tablefootmark{a} & Giz2002 \\
                        $ -16.7 \pm 1.5 $\tablefootmark{a} & Giz2002 \\
                        $ +30.1 \pm 1.5 $\tablefootmark{a} & Giz2002 \\
                        $ -5.7 $\tablefootmark{b} & Tor2006 \\
                        $ +43.74 \pm 0.11 $& Jef2018 \\
                        $ +5.74 \pm 0.15  $& Jef2018 \\
                        $ -16.08 \pm 0.22 $& Jef2018 \\
                        $ -18.89 \pm 0.12 $& Jef2018 \\
                        $ -17.14 \pm 0.35 $ & Jef2018 \\
                        $ -15.89 \pm 0.10 $ & Jef2018 \\
                   $ -12.56 \pm 0.10 $ & Jef2018 \\
                        $ 8.18 \pm 1.07 $ & Sch2019 \\
                        $ -0.4 \pm 1.0 $ & Spe2019 \\
                        $ -14.6 \pm 1.0 $ & Spe2019 \\
                \hline& 
        \end{tabular}
        \tablebib{
                Giz2002: \citet{Gizis2002};
                Tor2006: \citet{Torres2006};
                Jef2018: \citet{Jeffers2018};
                Sch2019: \citet{Schneider2019};
                Spe2019: \citet{Sperauskas2019};
        }
         \tablefoot{
                \tablefoottext{a}{
                        The associated error corresponds to the internal consistency menitoned in that work.
                }
                \tablefoottext{a}{
                        Mean value calculated from three different epochs. No uncertainty is provided.
                }
        }
\end{table}

\citet{Torres2006} first identified GJ\,1284 as a young object ($<$600\,Myr) based on several youth indicators.
\citet{Malo2013} proposed it as a member of either COL (99.2\%) or $\beta$PMG (99.9\%), depending on whether or not they included the mean RV value from \citet{Torres2006} in the analysis.
Following this work, GJ\,1284 has been commonly catalogued as a either a member of COL \citep{Malo2014a,Bell2015,Naud2017} or $\beta$PMG \citep{Elliott2016,Shkolnik2017}.
More recently, it has been rejected as a member of these two YMGs \citep{Gagne2018,Schneider2019} and was classified as a field star \citep{Janson2017} using the updated astrometry from {\it Gaia}.
Nevertheless,  the analysis of the membership of this object in a YMG is  based on the average value of the available RV measurements, not on its systemic velocity.

\citet{Malo2014a} compiled the value of several youth indicators for members of YMGs, including GJ\,1284.
In particular, four youth indicators were considered: the projected rotational velocity ($v\sin i$), the pseudo-equivalent width (pEW) of the H$\alpha$ emission line, the X-ray flux, and the pEW of the lithium 6707.8\,\text{\AA} absorption line. 
They report a value of $v\sin{i}=3.9\pm1.2\mathrm{km\,s^{-1}}$, in agreement with \citet{Torres2006}.
However, the authors note that this value might not be reliable because of the binarity of the system.
A value of $\mathrm{pEW(H\alpha)}=-3.3\,$\text{\AA} was obtained by \citet{Riaz2006} from medium-resolution spectra taken at the Cerro Tololo Inter-American Observatory (CTIO) 1.5 m telescope.
Again, this value could be affected by the binarity, although it also agrees with the $\mathrm{pEW(H\alpha)}=-3.23\pm0.08\,$\,\text{\AA}\ obtained by \citet{Jeffers2018} from high-resolution spectra.
The $\log{f_\mathrm{X}}=-11.44\,\mathrm{erg\,s^{-1}\,cm^{-2}}$ also comes from \citet{Riaz2006} using ROSAT data \citep{Voges1999} and the conversion relations described in \citet{Schmitt1995}.
Finally,  \citet{Torres2006} report that the Li 6708\text{\AA} line was not detected in their spectra.

\subsection{Wide companions} \label{WideCompanions}
Two other stellar systems are identified in the literature as common proper motion companions to GJ\,1284: GJ\,898 \citep{Shaya2011} and GJ\,897AB \citep{Kuiper1943}. We describe these two systems below.

\begin{figure}
        \resizebox{\hsize}{!}{\includegraphics{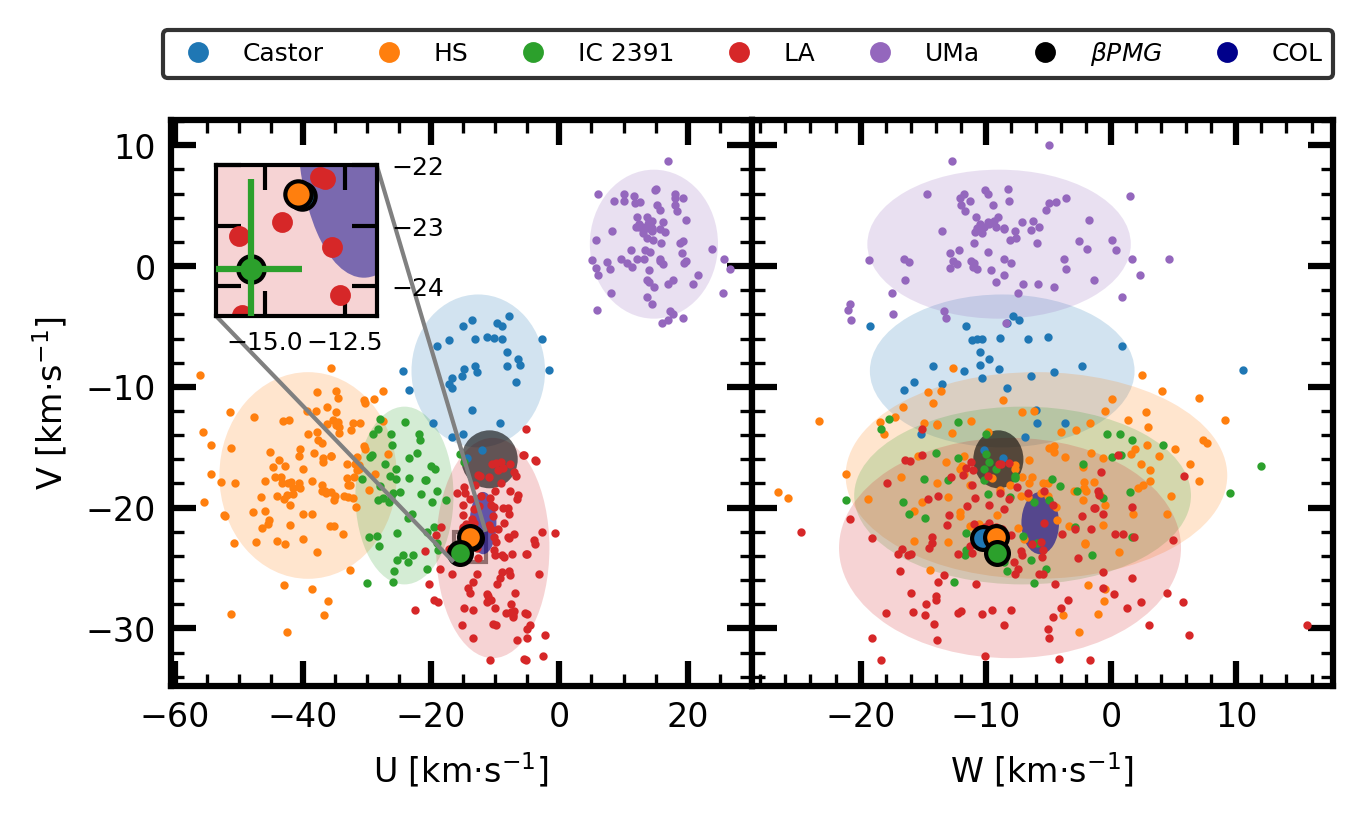}}
        \caption{Revised galactocentric UVW velocities of  GJ\,1284 (blue), GJ\,898 (orange), and GJ\,897AB (green) from this work compared to other YMG members from \citet{Montes2001}. The ellipses represent the 2$\sigma$ values of UVW for each YMG. Similarly, we overplot $\beta$PMG and COL as defined by \citet{Gagne2018}.}
        \label{UVWplot}
\end{figure}

GJ\,898 is a K6 \citep{Torres2006} located at a distance of $ d=14.549\pm0.006 $\,pc and classified as a member of the Local Association by \citet{Montes2001} according to its kinematics.
The projected separation in the sky to GJ\,1284 is 3.59 $ \deg $ or  $ \sim 2 \times 10^6$ \,au (roughly 1\,pc) at their mean distance.
The physical separation of 1.66\,pc together with a difference in their proper motions of over 10$ \sigma $ ({\it Gaia} eDR3; Table~\ref{stellarParamsTable}) may suggest this is a chance alignment.
Nonetheless, the galactocentric UVW velocities for GJ\,898 and GJ\,1284 (Sect.~\ref{youthSection})  are in very good agreement  (Table~\ref{stellarParamsTable}; Fig.~\ref{UVWplot}).
The difference in the astrometry can thus be explained as a projection effect consequence of their wide separation.
In Sect.~\ref{BondDiscussionSection} we analyse whether the two systems are physically bound.

GJ\,897AB was identified as a companion to GJ\,898 by \citet{Kuiper1943}, and therefore is also a companion to GJ\,1284.
It is a resolved tight visual binary \citep[$\rho=0.57\arcsec$; ][]{Mason2002} formed by two components of similar magnitude ($V_{A}=11.14\mathrm{\,mag}; V_{B}=11.17\mathrm{\,mag}$; WDS) and mass \citep[$ 0.5 \pm 0.1\,\mathrm{M_{\sun}}$][]{Heintz1986b}, which  suggests a similar spectral type M2--M3 \citep{Riaz2006,Torres2006}.
The orbit of the system is nearly edge-on ($i=89.30\degr$) with a period of 28.2 years \citep{Heintz1986a}.
While \citet{Woolley1970} list a distance for GJ\,897 of $ 14.5\pm2.7$\,pc and \citet{Gliese1991} give $ 12.9\pm9.9$\,pc, the separation of the components combined with the inclination of the orbit affects the quality of  the astrometric solutions of more modern catalogues.
For instance, Hipparcos lists a parallax of  $ 5 \pm 34 $\,mas \citep{VanLeeuwen2007}, while {\it Gaia} does not provide a full astrometric solution in the last eDR3.
In Table~\ref{stellarParamsTable} we list the parallax and proper motions from the dynamical solution provided by \citet{Tokovinin2018} for GJ\,897AB, which are in closest agreement with the values for GJ\,898. 
In this work we adopt for GJ\,897AB the distance and proper motions of GJ\,898 given their higher precision and the proximity of the two objects in the sky, which indicates they are most probably physical companions located at the same distance.
\citet{Makarov2008} also noted that the space velocities of GJ\,897AB and  GJ\,898  match those of the LA members from \citet{Montes2001}, and supported this assessment with other activity indicators (see  Sect.~\ref{youthSection}).

\begin{figure}
        \resizebox{\hsize}{!}{\includegraphics{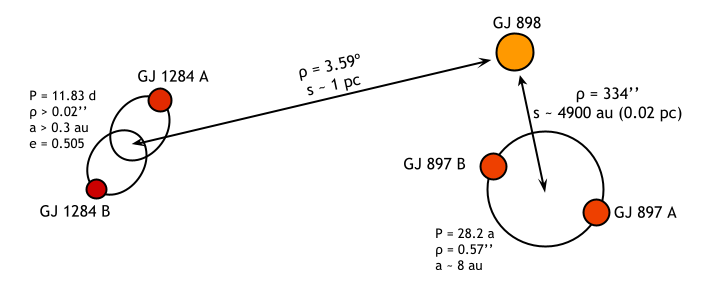}}
        \caption{Schematic view of the spatial configuration of the quintuple system made up of  GJ\,898, GJ\,1284AB, and GJ\,897AB. Sizes and distances are not to scale.}
        \label{systemDiagram}
\end{figure}

We performed a search for other common proper motion systems in a radius of up to 10 degrees using the {\it Gaia} eDR3 data, but it did not produce any additional candidates.
All together, GJ\,898, GJ\,897AB, and GJ\,1284 form a group of at least five stellar objects with very similar galactocentric velocities, with GJ\,898 being the most massive component.
A schematic of the configuration of this quintuple system is shown in Fig.~\ref{systemDiagram}.
The existence of this common proper motion system has an implication on the determination of the age of  GJ\,1284.
As previously mentioned, both GJ\,898 and GJ\,897AB have been associated with LA in the literature.
In Sect.~\ref{youthSection} we discuss the membership of the three systems in a YMG based on their kinematics and other youth indicators to infer an age for the quintuplet.

\begin{table*}
\caption{Stellar parameters of the GJ1284 system and its two wide companions.
}
\label{stellarParamsTable}
    \centering
    \begin{tabular}{l c c c c c}
    \hline\hline
    \noalign{\smallskip}
        Parameter & \multicolumn{4}{c}{Value} & References\tablefootmark{a} \\
    \noalign{\smallskip}
    \hline
    \noalign{\smallskip}
       &  \multicolumn{4}{c}{\textit{Name and identifiers}}  & \\
    \noalign{\smallskip}
        Main identifier    & G\,273-59      & HD\,221503        & \multicolumn{2}{c}{BD-17 6768}    &  \\
        Gl      & 1284          & 898   & 897A & 897B & \\
        TIC     & 9212046       &  434103018 & \multicolumn{2}{c}{434103039} & \\
         
    \noalign{\smallskip}
        & \multicolumn{4}{c}{\textit{Coordinates and kinematics}}  &\\
    \noalign{\smallskip}
        $\alpha$ (J2000) & 23:30:13.44  & 23:32:49.40    & 23:32:46.98 & 23:32:46.98  &  2 \\
        $\delta$  (J2000) & $ - $20:23:27.46    & $ - $16:50:44.32      &  $ - $16:45:12.34 & $ - $16:45:11.74    &  2 \\
        $\pi$ [mas]     & $ 62.67 \pm 0.09 $    & $ 68.70 \pm 0.06 $    & \multicolumn{2}{c}{$ 64.8 \pm 4.1 $\tablefootmark{b}}    &  2, 2, 3\\
        $d$ [pc]        & $ 15.906 \pm 0.019 $  & $14.549 \pm 0.006$ & \multicolumn{2}{c}{$15.15 \pm 0.46$} &  2, 2, 3 \\
        $\mu_{\alpha}\cos{\delta}$ [mas\,a$^{-1}$]  & $313.50 \pm 0.15$ & $340.98 \pm 0.12$       & \multicolumn{2}{c}{$352.0 \pm 2.6$}   & 2, 2, 3  \\
        $\mu_{\delta}$ [mas\,a$^{-1}$] & $-205.41 \pm 0.15$     & $-218.98 \pm 0.07$       & \multicolumn{2}{c}{$-216.1 \pm 2.7$} &  2, 2, 3 \\
        V$_\mathrm{r}$ [km\,s$^{-1}$]  & $0.84 \pm 0.14$     & $-0.845 \pm 0.0012$ & \multicolumn{2}{c}{$-1.6 \pm 1.7$} & 1,  2,  4 \\
        U [km\,s$ ^{-1} $]         &  $-13.837 \pm 0.026$       & $-13.942 \pm 0.009$      & \multicolumn{2}{c}{$-15.4 \pm 0.8$}   &  1 \\
        V [km\,s$ ^{-1} $]         & $-22.499 \pm 0.021$        & $-22.471 \pm 0.007$      & \multicolumn{2}{c}{$-23.7 \pm 0.7$    }       &  1 \\
        W [km\,s$ ^{-1} $]        & $-10.231 \pm 0.015$    & $-9.184 \pm 0.003$          & \multicolumn{2}{c}{$-9.1 \pm 1.6$}              & 1 \\
   
    \noalign{\smallskip}
         & \multicolumn{4}{c}{\textit{Spectral type and photometry}} & \\
    \noalign{\smallskip}
        SpT        & M2--M2.5V\,+\,M3--M3.5V            & K6.0V & \multicolumn{2}{c}{M2.5V}     &  1, 4, 5  \\ 
        $B$ [mag]    & $ 12.59 \pm 0.03 $       & $ 9.890 \pm 0.019 $   & \multicolumn{2}{c}{$ 11.79 \pm 0.03 $}  & 6, 7, 6 \\
        $V$ [mag]    & $11.11 \pm 0.03$ & $8.600 \pm 0.013$     & \multicolumn{2}{c}{$10.34 \pm 0.04$}      &  6, 7, 6 \\
        $G$ [mag]    & $9.9210 \pm 0.0020$    & $8.0816 \pm 0.0008$     & $10.033 \pm 0.005$ & $10.058 \pm 0.004$ & 2 \\
        $BP$ [mag]    & $ 11.319 \pm 0.003$    & $ 8.829 \pm 0.004 $    & $10.548 \pm 0.008$ & $10.556 \pm 0.007$ & 2 \\
        $RP$ [mag]    & $ 8.777 \pm  0.004 $    & $ 7.251 \pm 0.004 $   & $8.198 \pm 0.005$ & $8.210 \pm 0.009$ & 2 \\
        $J$ [mag]    & $7.200 \pm 0.019$        & $6.236 \pm 0.019$     & \multicolumn{2}{c}{$6.71 \pm 0.20$}     & 8 \\
        $H$ [mag]    & $6.61 \pm 0.04$      & $5.61 \pm 0.03$   & \multicolumn{2}{c}{$6.08 \pm 0.03$}      &  8\\
        $K$ [mag]    & $6.329 \pm 0.026$      & $5.473 \pm 0.016$       & \multicolumn{2}{c}{$ 5.858 \pm 0.016$}  & 8 \\
        $g$ [mag]    & $12.4084 \pm 0.0010$      & $10.86 \pm 0.05$             & \multicolumn{2}{c}{$9.70 \pm 0.04$}&  9 \\
        $r$ [mag]    & $10.50 \pm 0.01$      & $8.308 \pm 0.011$                & \multicolumn{2}{c}{$9.91 \pm 0.11$      }&  9 \\
        $i$ [mag]    & $10.88 \pm 0.05$       & $9.72 \pm 0.08$         & \multicolumn{2}{c}{$8.765 \pm 0.021$}   &  9 \\
        $NUV$ [mag]    & $18.02 \pm 0.04$       & $16.900 \pm 0.015$            & \multicolumn{2}{c}{$17.253 \pm  0.017$} &  10, 11, 10\\
    \noalign{\smallskip}
        & \multicolumn{4}{c}{\textit{Additional information}}  &\\
    \noalign{\smallskip}
        $v\sin{i}\mathrm{\ [km\,s^{-1}]}$    & $3.9 \pm 1.2$ & $7.47 \pm 1.00$ & \multicolumn{2}{c}{$7.1 \pm 1.2$} & 12, 13, 4 \\
        $\mathrm{pEW(H\alpha)} $\,[\text{\AA}]     & $-3.23 \pm 0.08$  & $0.12 \pm 0.04$ &  \multicolumn{2}{c}{$-1.051$} &  12, 13, 14 \\
        $\log{L_\mathrm{X}}\mathrm{\ [erg\,s^{-1}]}$\tablefootmark{c}  &  $29.04 \pm 0.04$ & $28.15 \pm 0.11$ & \multicolumn{2}{c}{$29.22 \pm 0.044$} & 15 \\
        $P_\mathrm{rot}$ [d]                    &       $ 7.45 \pm 0.1 $ &       $ 11.5 \pm  0.1 $ &     \multicolumn{2}{c}{$ 5.00 \pm 0.02 $} & 1 \\
    \hline
    \end{tabular}
    \tablebib{
        (1) This work;
        (2) {\it Gaia} eDR3: \citet{Prusti2016, Brown2020};
        (3) \citet{Tokovinin2018};
        (4) \citet{Torres2006};
        (5) \citet{Riaz2006} ;
        (6) APASS: \citet{Henden2016}; 
        (7) SKY2000: \citet{Myers2015};
        (8) 2MASS: \citet{Skrutskie2006};
        (9) UCAC4: \citet{Zacharias2013};
        (10) \citet{Schneider2019};
        (11)  \citet{Riedel2017};
        (12) \citet{Jeffers2018};
        (13) \citet{Lopez-Santiago2010};
        (14) \citet{Jones2016};
        (15) ROSAT: \citet{Voges1999}.
    }
    \tablefoot{
        \tablefoottext{a}{
                Reference numbers for the three systems are separated by commas; a single number is provided whenever it is the same for all of them.
        }
        \tablefoottext{b}{
                For the purposes of this work we adopt the parallax value from GJ\,898 given the high uncertainty on the parallax determination of GJ\,897 and their proximity and shared kinematics.
        }
        \tablefoottext{c}{
                Calculated using the relations from \citet{Riaz2006}.
        }
    }
\end{table*}


\section{Observations}\label{ObservationsSection}

\begin{table*}
\caption{Instrumental configuration of the spectrographs used in this paper.}
\label{specParamTable}
    \centering
    \begin{tabular}{l l c c c c c c l}
    \hline\hline
    \noalign{\smallskip}
         Instrument & Telescope & R & No. orders & $\Delta\lambda$ & No. epochs & $\Delta$t & t$ _{\mathrm{exp}}$  & Configuration \\
          & & & & [nm]& & [d] & [s]& \\
    \noalign{\smallskip}
    \hline
    \noalign{\smallskip}
        HERMES      & Mercator          & 85\,000 & 54 & 377--900 & 11 & 43& 1\,800 & HRF \\
        CAFE        & Calar Alto 2.2m   & 62\,000 & 84 & 396--950 & 7 & 7 & 1\,800 & \\
        HARPS-N     & Telescopio Nazionale Galileo & 115\,000 &69 & 383--690 & 4 & 4& 900 & GIARPS \\
        NRES        & LCOGT 1-metre telescopes            & 53\,000  & 67 & 380--870 & 5 & 4 & 1\,800 & \\
    \hline& & 
    \end{tabular}
\end{table*}{}

\begin{figure}
  \resizebox{\hsize}{!}{\includegraphics{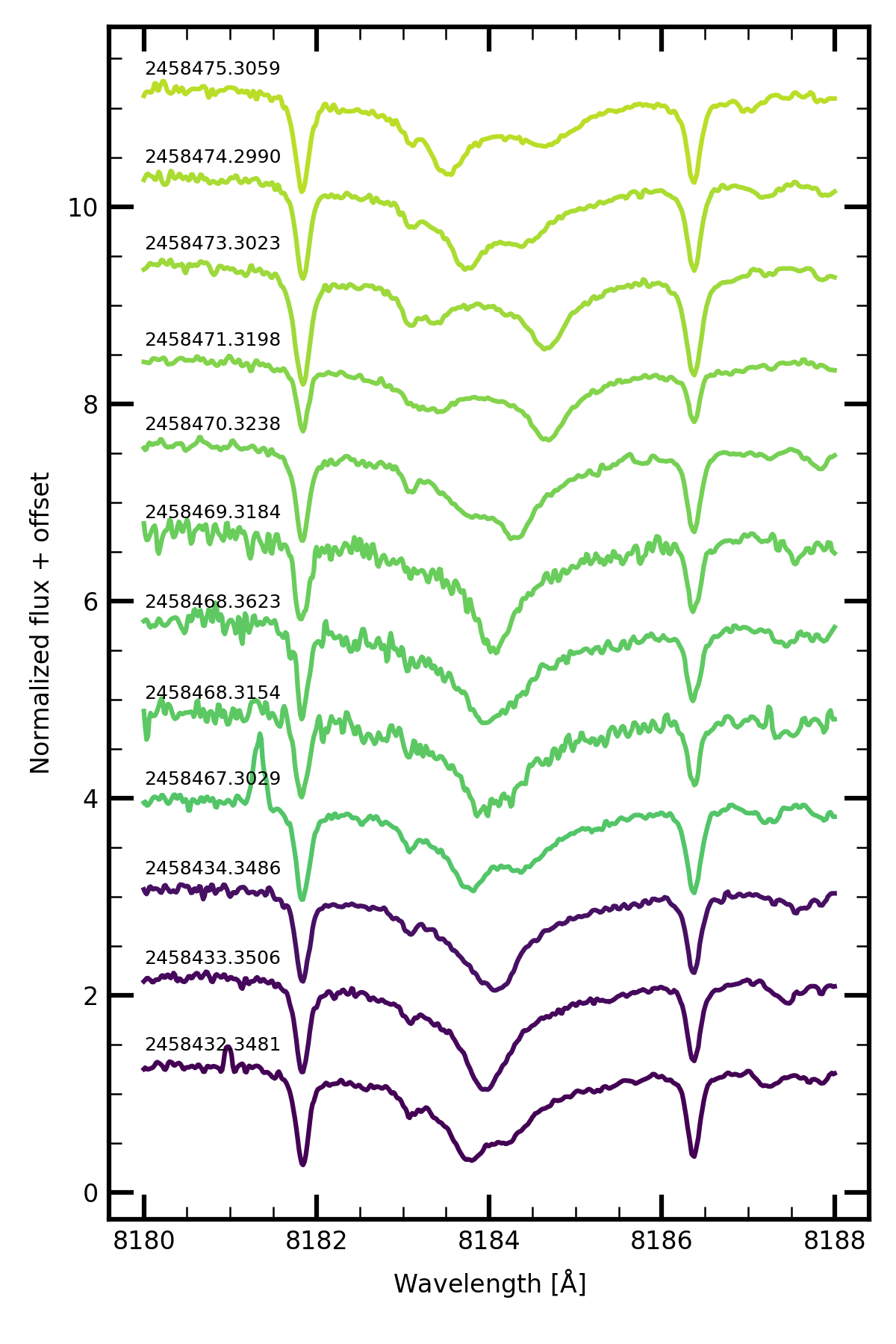}}
  \caption{HERMES spectra of GJ\,1284 showing the \ion{Na}{i} $ \lambda\, 8183.25 $\,\text{\AA} line.}
  \label{HERMESspectra}
\end{figure}

\begin{figure}
        \resizebox{\hsize}{!}{\includegraphics{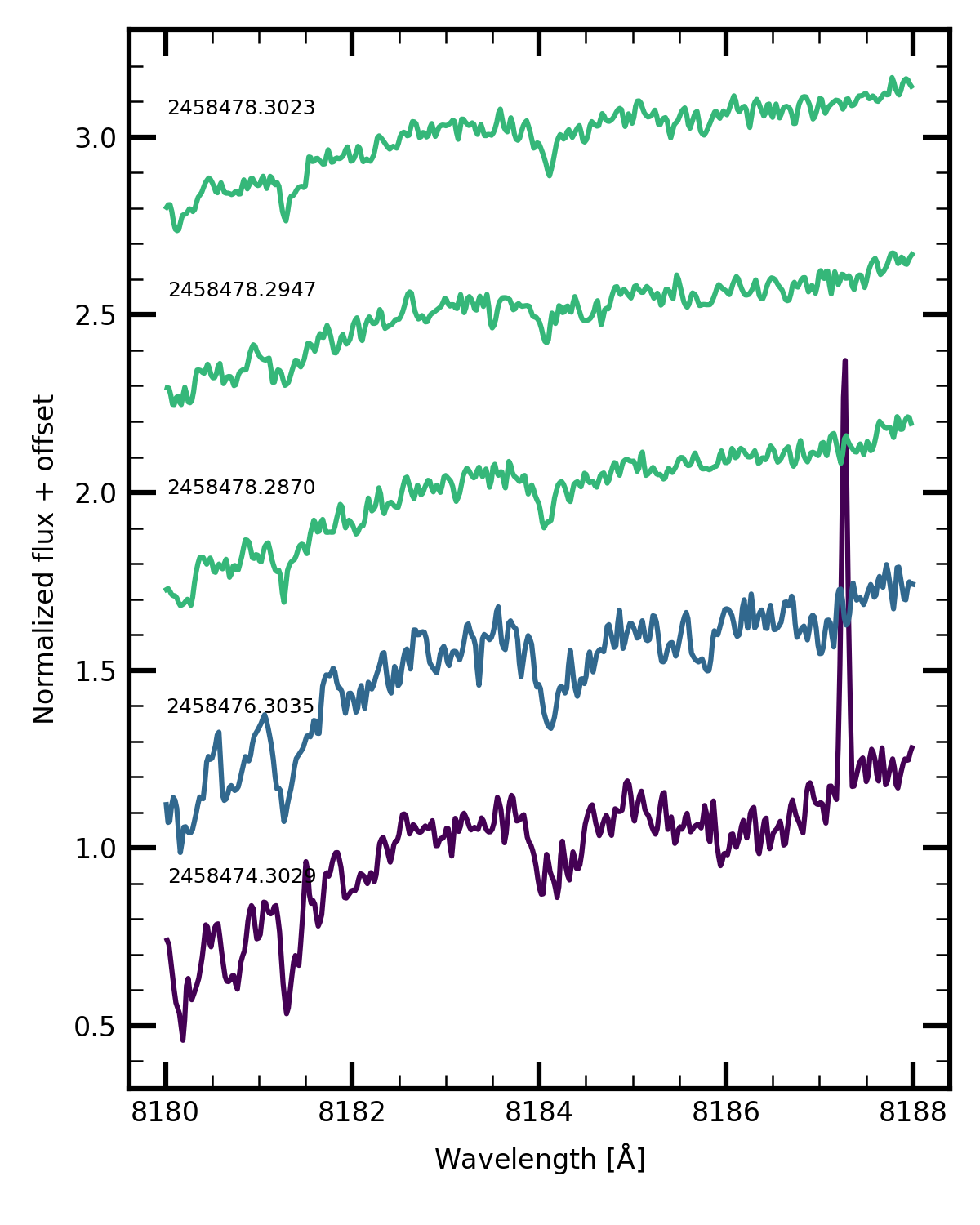}}
        \caption{NRES spectra of GJ\,1284 centred around the \ion{Na}{i} $ \lambda\, 8183.25 $\,\text{\AA} line.}
        \label{NRESspectra}
\end{figure}

\begin{figure}
  \resizebox{\hsize}{!}{\includegraphics{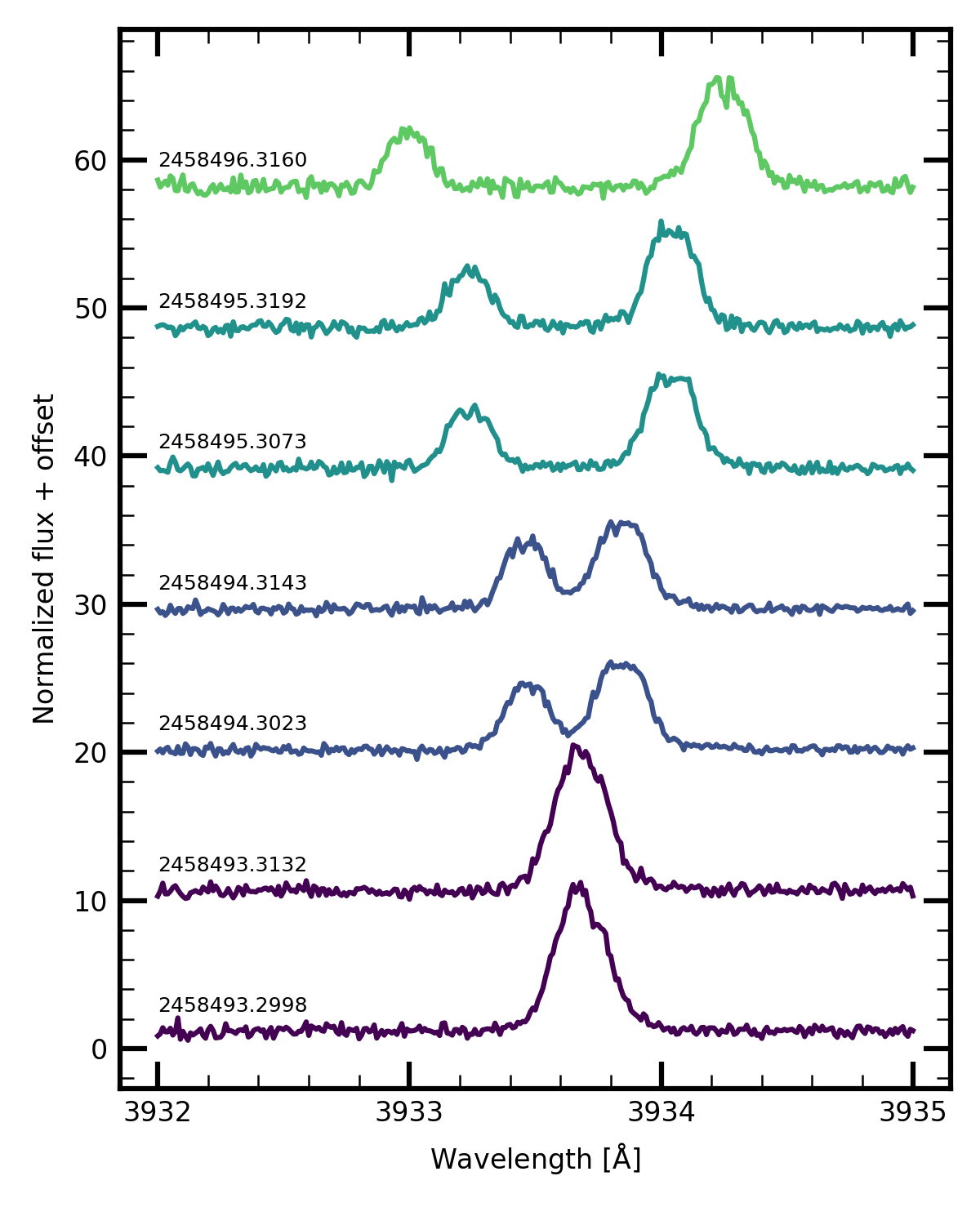}}
  \caption{HARPS-N spectra of GJ\,1284 focused on the \ion{Ca}{ii}-H $ \lambda\, 3968.47 $\,\text{\AA} line.}
  \label{HARPSspectra}
\end{figure}

\begin{figure}
        \resizebox{\hsize}{!}{\includegraphics{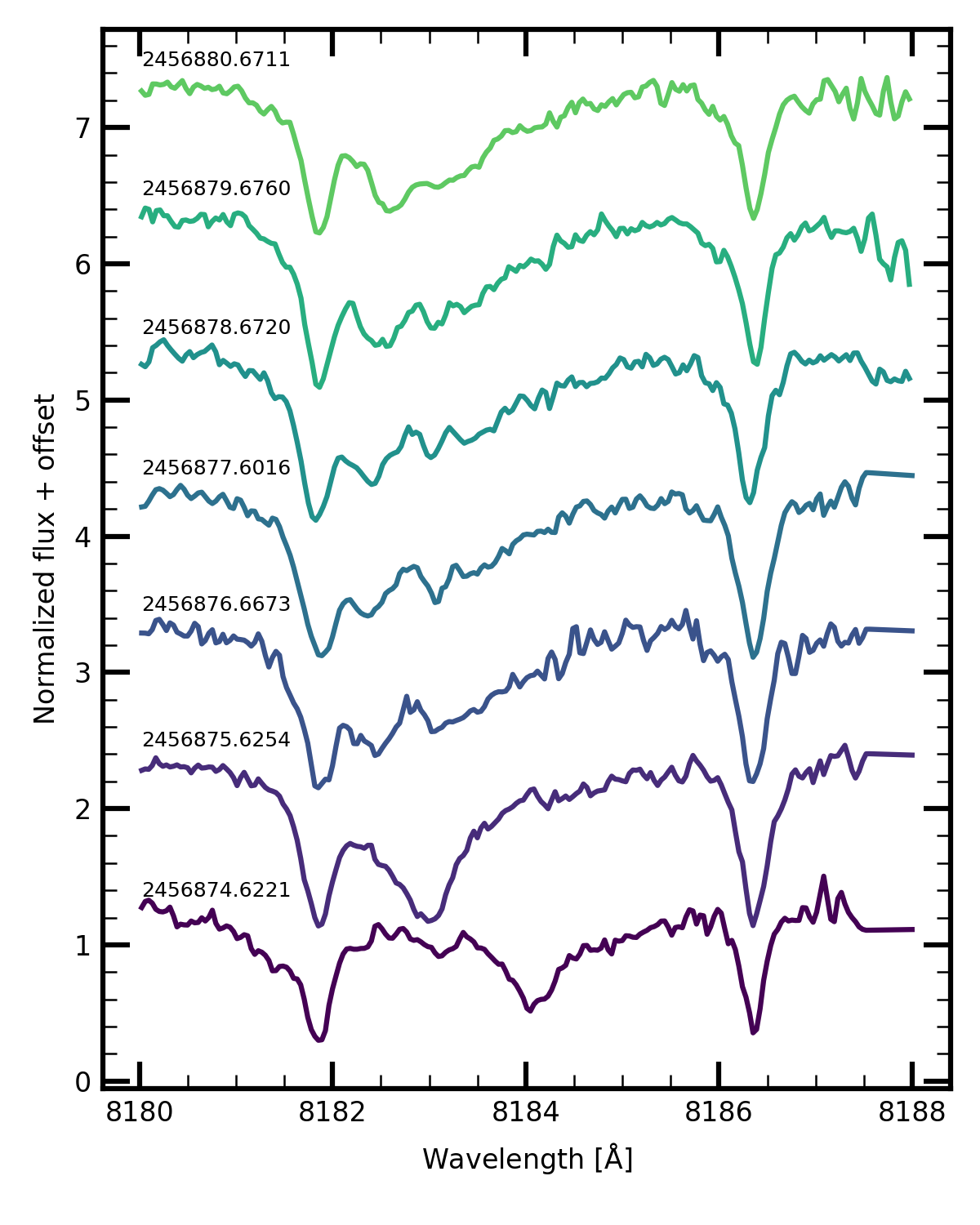}}
        \caption{CAFE spectra of GJ\,1284 depicting the \ion{Na}{i} $ \lambda\, 8183.25 $\,\text{\AA} line.}
        
        \label{CAFEspectra}
\end{figure}

We gathered 27 spectra for GJ\,1284 with four distinct high-resolution spectrographs (Table~\ref{specParamTable}).
Additionally,  this object and the two wide companions described in the previous section were  observed by TESS, although the individual components of GJ\,1284 and GJ\,897AB are not resolved in the images.

\subsection{Spectroscopy}
\subsubsection{HERMES}
We first observed and identified GJ\,1284 as an SB2 with the High Efficiency and Resolution Mercator Echelle Spectrograph \citep[HERMES;][]{Raskin2011} as part of observing programme CAT-141-MULTIPLE-2/18B (PI: Cardona Guill\'en).
We also obtained additional spectra with this same instrument through progammes 128-Mercator4/18B (PI: Montes) and  ITP18\_8 (PI: Tremblay).
HERMES is a fibre-fed echelle  spectrograph mounted on the Mercator 1.2 m semi-robotic telescope located in the Roque de los Muchachos Observatory (ORM) on La Palma, Spain. 
A total of 11 spectra were gathered over several runs spanning from 9 November to 22 December 2018 under clear conditions and a seeing of $\sim 0.8$\,arcsec.
The spectra were taken using the high-resolution mode (HRF) with an exposure time of 1\,800\,s. 
This mode provides a spectral resolution R$\sim$85\,000 and a wavelength coverage from 377 to 900\,nm through a 2.5\,arcsec fibre equipped with an image slicer. 
The data were reduced by the automated pipeline and RV toolkit (HermesDRS\footnote{\url{http://www.mercator.iac.es/instruments/hermes/drs/}}). 
This tool provides  one-dimensional wavelength-calibrated spectra using the standard reduction procedure (bias subtraction, flat-field correction, order extraction, and wavelength calibration) from which we derive the RV measurements described in Sect.~\ref{RVDetermSection}.
Figure~\ref{HERMESspectra} shows the HERMES spectra for each epoch zoomed in around the \ion{Na}{i} $ \lambda\, 8183.25 $\,\text{\AA}.

\subsubsection{NRES}
We obtained five additional spectra on three separate nights (21, 23, and 25 December 2018) using the Network of Robotic Echelle Spectrographs \citep[NRES;][]{Siverd2018} as part of programme KEY2017AB-002a (PI: Brown).
NRES is a set of identical high-resolution spectrographs mounted on the 1 m telescopes of the Las Cumbres Observatory network (LCOGT; \citealt{Brown2013}).
Each spectrograph is fibre-fed by two telescopes simultaneously, and has a resolution R$\sim$53\,000 and a wavelength coverage from 380 to 860\,nm. 
The spectra used in this work were obtained using the node located in the South African Astronomical Observatory (SAAO) in Sutherland, South Africa. 
Currently, three other NRES elements are available in the LCOGT nodes located in Chile, the United States, and Israel.
The data were reduced using the automatic NRES pipeline\footnote{\url{https://lco.global/documentation/data/nres-pipeline/}}.
The pipeline yields a one-dimensional spectrum calibrated in wavelength similarly to HERMES.
In Fig.~\ref{NRESspectra} we plot the NRES spectra for each epoch showing the \ion{Na}{i} $ \lambda\, 8183.25 $\,\text{\AA}.

\subsubsection{HARPS-N}
We also observed the target using the High Accuracy Radial velocity Planet Searcher North \citep[HARPS-N;][]{Cosentino2012} on four consecutive nights between 9 and 12 January 2019 under programme CAT-141-MULTIPLE-2/18B (PI: Cardona Guill\'en).
HARPS-N is an echelle spectrograph mounted on the 3.58 m Telescopio Nazionale Galileo (TNG) also located in the ORM.
The design of the instrument is similar to its predecessor HARPS \citep{Mayor2003}, covering a wavelength range from 383 to 690 nm with a resolution of R$\sim$115\,000.
We obtained two spectra with exposure times of 900\,s on each of the observing nights, except on  12 January 2019 when we obtained only one spectrum.
These spectra were reduced automatically at the end of each night using the standard HARPS-N data reduction pipeline \citep{Cosentino2014,Smareglia2014}.
Figure ~\ref{HARPSspectra} shows the HARPS-N spectra around the \ion{Ca}{ii}-H $ \lambda\, 3968.47 $\,\text{\AA} line.

\subsubsection{CAFE}
We also used spectra collected by \citet{Jeffers2018} with the Calar Alto Fiber-fed Echelle spectrograph \citep[CAFE;][]{Aceituno2013}.
CAFE is a fibre-fed echelle spectrograph mounted on the Calar Alto 2.2 m telescope in Centro Astronómico Hispano Alemán (CAHA) in Almería, Spain.
The spectrograph has a wavelength coverage ranging from 396 to 950\,nm spanning 84 orders with a resolution R$\sim$62\,000.
A total of seven spectra were taken over the same number of consecutive nights from 5 August to 11 August 2014 with exposure times of 1\,800\,s.
The data were reduced using a version of the REDUCE package \citep{Piskunov2002}.
More details about the observations and the data reduction process can be found in \citet{Jeffers2018}.
Figure~\ref{CAFEspectra} shows the CAFE spectra at each epoch around the \ion{Na}{i} $ \lambda\, 8183.25 $\,\text{\AA}.

\subsection{Photometry} \label{TESS}
The Transiting Exoplanet Survey Satellite (TESS; \citealt{Ricker2014}) is a space mission launched on 18 April 2018. 
During its two-year prime mission, which started on 25 July 2018, TESS   monitored more than 200\,000 main-sequence stars, covering almost all of the sky using its four cameras with a combined field of view of $24\degr\times96\degr$ in continuous campaigns of 27 days. 
A pre-selection of targets were monitored with a two-minute cadence.

\begin{figure*}
        \centering
        \includegraphics[width=17cm]{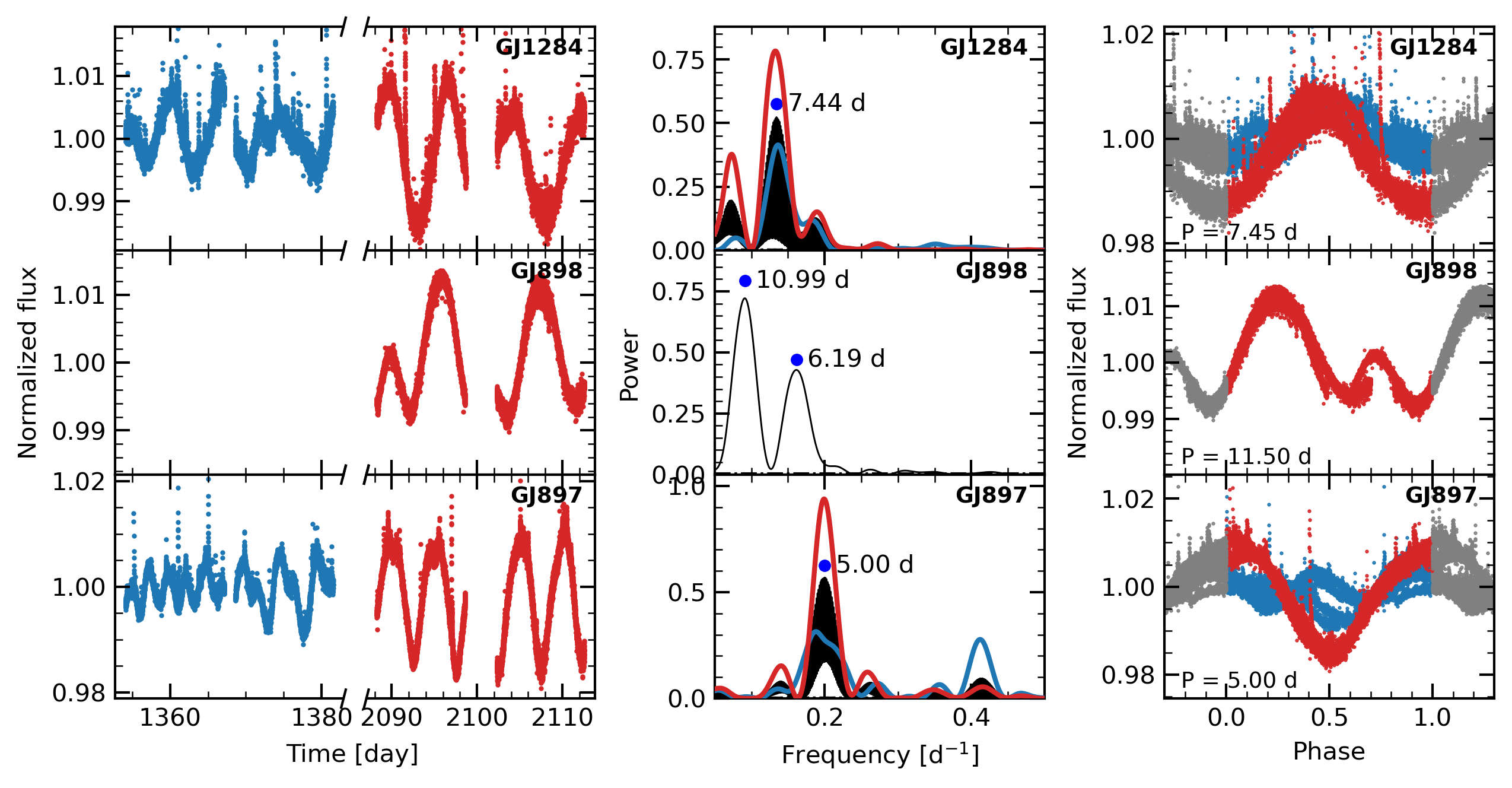}
        \caption{TESS light curves and periodograms of GJ\,1284 (top), GJ\,898 (middle)  and GJ\,897 (bottom).
        		\textit{Left}: TESS photometric time series.
                \textit{Middle}: GLS periodogram of the photometric time series.
                \textit{Right}: Light curve phase-folded with the selected rotation periods in Table~\ref{stellarParamsTable}.
                We adjust the ordinate axis range to crop the stellar flares to better visualise the variations in the light curve due to the stellar rotation.
                The colours in all the panels correspond to data from Sector 2 (blue) or Sector 29 (red) or both combined (black).
        }
        \label{TESSfigure}
\end{figure*}

GJ\,1284 and its two co-moving companions GJ\,898 and GJ\,897AB were observed by TESS as part of Sector 2, from 22 August to 20 September 2018, and Sector 29, from 26 August to 22 September 2020.
For GJ\,1284 and GJ\,897AB there is a two-minute cadence light curve available for each of these sectors, while for GJ\,898 only the light curve  for Sector 29 can be retrieved.
The target input catalogue (TIC) numbers for the three objects are shown in Table~\ref{stellarParamsTable};  their light curves are publicly available and can be downloaded from the Mikulski Archive for Space Telescopes (MAST) Portal \footnote{\url{https://mast.stsci.edu/portal/Mashup/Clients/Mast/Portal.html}}.
The two-minute-cadence photometry includes two different sets of data: the single aperture photometry (SAP) light curve,  and the pre-search data conditioning (PDC) light curve.
The PDC light curve attempts to correct the common instrumental effects present in the SAP light curve.
In this work we use the SAP light curve to avoid corrections that may affect the activity-induced photometric variations, especially in relatively slow rotators with low-amplitude variability like ours.
The two-minute cadence light curves of the three systems are shown in  Fig.~\ref{TESSfigure}.

\section{Data analysis} \label{DataAnalysisSection}
\subsection{Radial velocities determination\label{RVDetermSection}}
The usual approach to calculating the RV of an object with the cross-correlation function (CCF) technique can lead to the determination of less precise values in SBs because of the blending of the spectral lines of the binary components.
For this reason we used \texttt{TODMOR} \citep{Zucker2004} to derive the RV values from all the spectra gathered for this object.
\texttt{TODMOR} is a multi-order implementation of the TwO-Dimensional CORrelation technique (\texttt{TODCOR}; \citealt{Zucker1994}). 
It uses a combination of two template spectra to simultaneously calculate the CCF for both components of the binary system.
This technique is applied to each echelle order, combining the resulting CCFs into a single one to derive the RV\@.

We used a set of template spectra created using PHOENIX stellar models \citep{Husser2013} as a reference to calculate the CCF. 
These templates have varying values of effective temperature ($ T_\mathrm{eff} $) that can be broadened to match the projected rotational velocity ($ v\sin{i} $) of the star.
We fixed the value for the metallicity $ [\mathrm{Fe/H}]=0 $ given the proximity of the system to the Sun.
We used a range for  $ T_\mathrm{eff} $ from 3000 to 4000\,K in steps of 100\,K and for $ v\sin{i} $ from 0 to 10 $ \mathrm{\,km\,s^{-1}} $ in steps of 1 $ \mathrm{\,km\,s^{-1}} $.
An additional parameter $\alpha$ was included to account for the flux-ratio between the two template spectra to calculate the CCF.
We calculated the values that provide the best fit of the CCF using the HERMES spectrum observed closer to quadrature, when the lines of the binary components are most separated.
We found that the optimal parameter values for the stellar templates are $T_\mathrm{eff}=3800\pm100\,\mathrm{K}$ and $v\sin{i}=6\pm1\mathrm{\,km\,s^{-1}}$ for the primary component and $ T_\mathrm{eff}=3600\pm100\,\mathrm{K}$ and $v\sin{i}=8\pm1\mathrm{\,km\,s^{-1}}$ for the secondary, with a flux ratio $\alpha=0.47\pm0.10$.
These templates are used to derive the RV for the two components simultaneously at each epoch \citep[see][for more details]{Zucker1994}.
Taking advantage of the CCF per spectral order provided by \texttt{TODMOR}, we also inspected and excluded from the analysis those orders that are heavily contaminated by telluric lines, and those with low signal-to-noise ratios (S/N), mainly in the blue part of the spectrum where the peaks of the two components are not apparent in the CCF.
The final RVs derived by \texttt{TODMOR} are given in Table~\ref{finalRVsTable}.

In addition, we calculated and applied the barycentric correction (BVCOR) to each RV epoch.
HARPS-N  applies this correction to the spectra themselves so that the extracted RV values are already in the barycentric reference frame. Moreover, it also provides the Barycentric Julian Day (BJD).
For HERMES the value for the BVCOR and the BJD of each exposure is calculated as part of the pipeline and provided in the header of each image.
Instead, CAFE and NRES do not provide these values as part of the automatic pipeline.
For these instruments, we calculated the BJD and BVCOR at each epoch with the routines included in the \texttt{astropy} Python package\footnote{\url{https://docs.astropy.org/en/stable/index.html}}.
This package uses an implementation of the expressions in \citet{Wright2014}.

To improve the determination of the RVs we also accounted for relative drift in the wavelength calibration of the different instruments. 
To palliate this issue we tracked the position of the telluric lines at each epoch using the IRAF routine \texttt{fxcor} \citep{Tody1986} to calculate a one-dimensional CCF.
We correlated the HERMES spectrum with the highest S/N, which provides the widest wavelength coverage of all the spectrographs, against the rest of the spectra and chose  the regions where the presence of the telluric lines is more apparent (e.g. 8119--8179, 8203--8324, 8928--8997\,\text{\AA}).
The drift of the telluric lines relative to the reference spectrum is calculated from the highest peak of the resulting CCF.
For the HERMES spectra we find that  the internal drift is comparable ($<0.5\mathrm{\,km\,s^{-1}}$) to the uncertainty associated with the individual RV calculated using \texttt{TODMOR}.
In the case of the CAFE spectra the internal drift is of the same order as HERMES, but there is a drift of $ \sim1\mathrm{\,km\,s^{-1}} $ between the two instruments.
For NRES the internal drift appears to be about $ \sim 0.5\mathrm{\,km\,s^{-1}} $, similar to HERMES and also comparable to the systematic drift between the two instruments.
Lastly, the internal drift in the HARPS-N spectra is lower than the precision we can achieve with this method of $0.1\mathrm{\,km\,s^{-1}}$ (i.e. there is no noticeable internal drift).
 Meanwhile, there seems to be a systematic drift of  $ \sim -0.3\mathrm{\,km\,s^{-1}} $ with respect to the reference HERMES spectrum, lower than the internal drift of this spectrograph.
For the final analysis, we corrected the RV values derived with \texttt{TODMOR} from the drifts measured using the telluric lines at each epoch (Table~\ref{finalRVsTable}).
The internal drifts show that the stability of the four spectrographs is good enough to characterise our binary system, which presents variations of over $60\mathrm{\,km\,s^{-1}}$.

\begin{figure}
  \resizebox{\hsize}{!}{\includegraphics{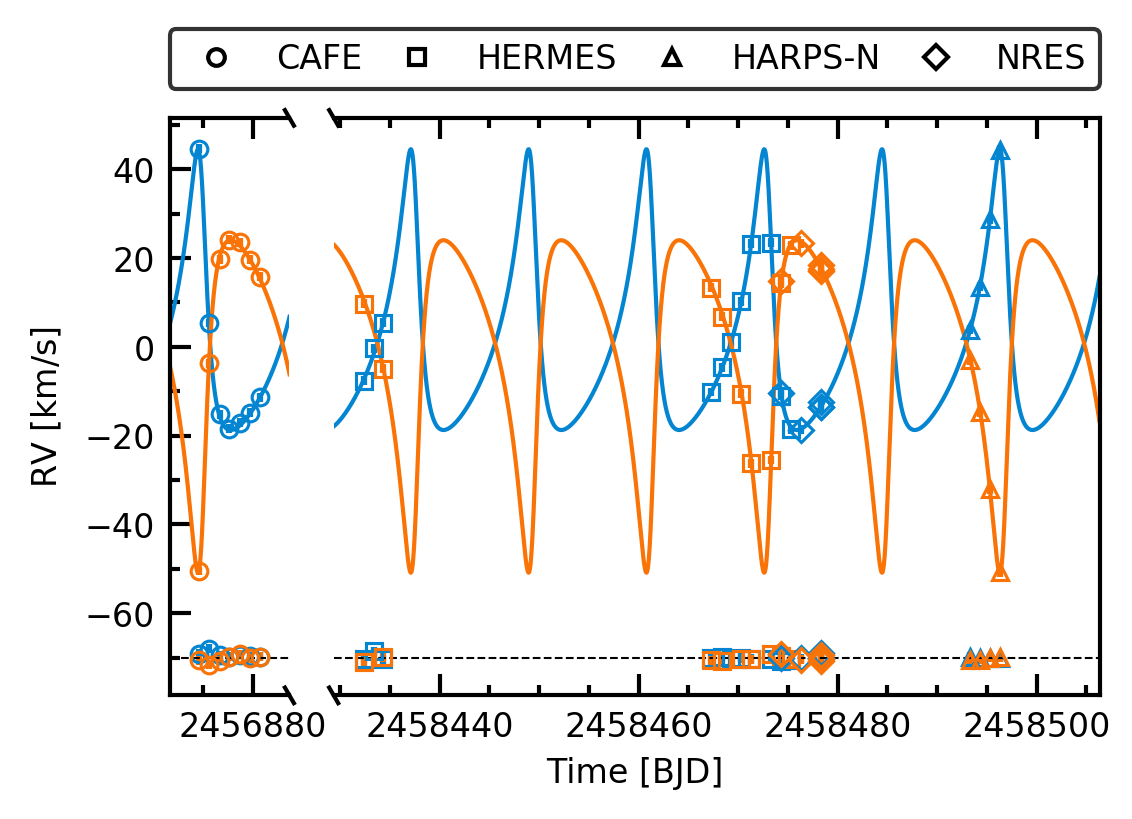}}
  \resizebox{\hsize}{!}{\includegraphics{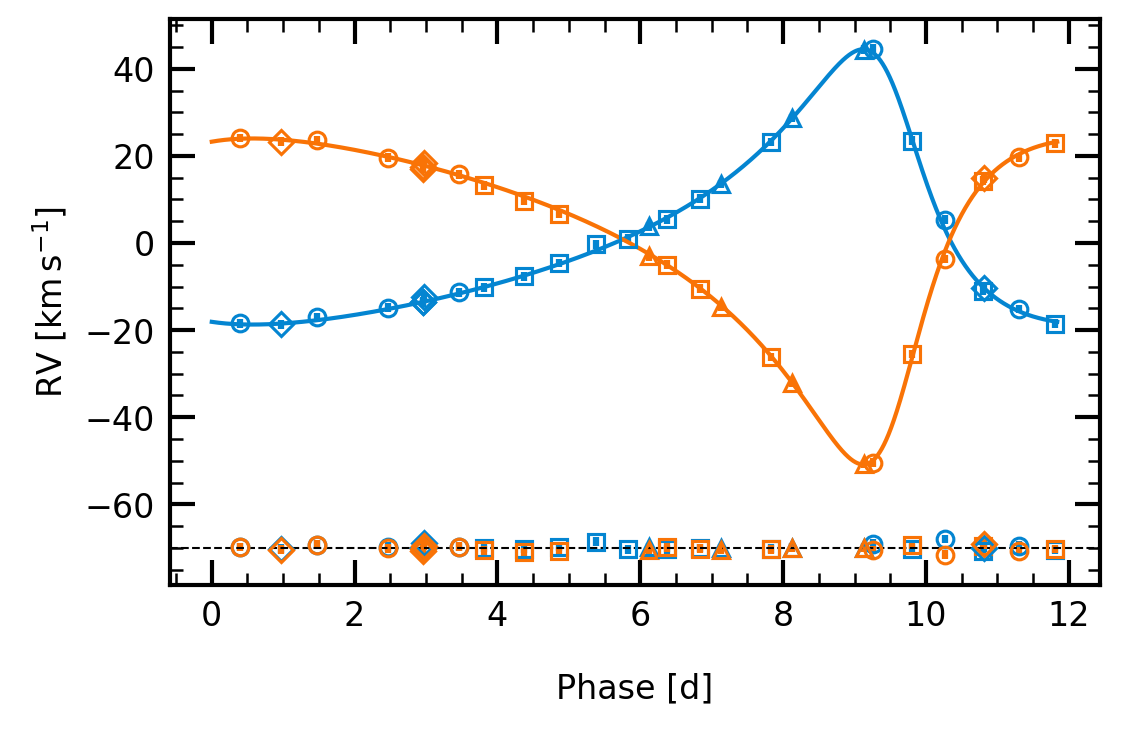}}
  \caption{RV values for the four different spectrographs (CAFE, HERMES, HARPS-N, and NRES) and orbital solution for the primary (blue) and secondary (orange) component of GJ\,1284 as a function of time (\textit{top panel}) and orbital phase (\textit{bottom panel}).
  The residuals of the Keplerian fit calculated with \texttt{rvfit} are shown at the bottom of each panel using the same scale.
  }
  \label{orbitFigure}
\end{figure}

Table~\ref{finalRVsTable} includes the final RVs used in the orbital parameter determination (Sect.~\ref{orbitParametersSection}) and their uncertainties together with the BJD, barycentric, and telluric correction at each epoch.
In Fig.~\ref{orbitFigure} we plot the RVs as a function of time (top panel) and orbital phase (bottom panel).

\begin{table*}
\centering
\caption{Summary of the RV measurements and the associated uncertainties obtained from the spectra along the BJD and barycentric corrections at each epoch.
}
\label{finalRVsTable}
\begin{tabular}{cccccccl}
         \hline\hline
        \noalign{\smallskip}
        BJD &        BVCOR &     RV$ _\mathrm{tell} $ &      RV$ _1 $ &     $ \sigma_\mathrm{RV1} $ &      RV$ _2 $ &     $ \sigma_\mathrm{RV2} $ &    Inst \\\relax
        [d] & [$ \mathrm{\,km\,s^{-1}}$] &[ $ \mathrm{\,km\,s^{-1}}$] & [$ \mathrm{\,km\,s^{-1}}$] & [$ \mathrm{\,km\,s^{-1}}$] & [$ \mathrm{\,km\,s^{-1}}$ ]& [$ \mathrm{\,km\,s^{-1}}$]  &   \\
        \hline
        \noalign{\smallskip}
        $ 2456874.6254 $ &  $ 14.97 $ &  $ 1.24 $ &  $ 44.65 $ & $ 0.20 $ & $ -50.47 $ & $ 0.21 $ &    CAFE \\
        $ 2456875.6288 $ &  $ 14.55 $ &  $ 1.44 $ &   $ 5.36 $ & $ 0.18 $ &  $ -3.67 $ & $ 0.19 $ &    CAFE \\
        $ 2456876.6708 $ &  $ 14.03 $ &  $ 1.44 $ & $ -15.17 $ & $ 0.23 $ &  $ 19.70 $ & $ 0.22 $ &    CAFE \\
        $ 2456877.6051 $ &  $ 13.77 $ &  $ 1.55 $ & $ -18.48 $ & $ 0.22 $ &  $ 24.14 $ & $ 0.22 $ &    CAFE \\
        $ 2456878.6756 $ &  $ 13.17 $ &  $ 1.99 $ & $ -17.04 $ & $ 0.43 $ &  $ 23.56 $ & $ 0.52 $ &    CAFE \\
        $ 2456879.6797 $ &  $ 12.74 $ &  $ 1.22 $ & $ -14.85 $ & $ 0.49 $ &  $ 19.62 $ & $ 0.49 $ &    CAFE \\
        $ 2456880.6748 $ &  $ 12.32 $ &  $ 1.22 $ & $ -11.36 $ & $ 0.47 $ &  $ 15.78 $ & $ 0.54 $ &    CAFE \\
        $ 2458432.3611 $ & $ -25.72 $ & $ -0.04 $ &  $ -7.67 $ & $ 0.09 $ &   $ 9.75 $ & $ 0.14 $ &  HERMES \\
        $ 2458433.3635 $ & $ -25.97 $ & $ -0.74 $ &  $ -0.33 $ & $ 0.16 $ &      ... & ... &  HERMES \\
        $ 2458434.3614 $ & $ -26.19 $ &  $ 0.00 $ &   $ 5.45 $ & $ 0.10 $ &  $ -4.97 $ & $ 0.12 $ &  HERMES \\
        $ 2458467.3126 $ & $ -28.90 $ & $ -0.13 $ & $ -10.16 $ & $ 0.09 $ &  $ 13.27 $ & $ 0.13 $ &  HERMES \\
        $ 2458468.3350 $ & $ -28.96 $ & $ -0.31 $ &  $ -4.58 $ & $ 0.10 $ &   $ 6.76 $ & $ 0.14 $ &  HERMES \\
        $ 2458468.3719 $ & $ -28.96 $ & $ -0.31 $ &  $ -4.58 $ & $ 0.10 $ &   $ 6.76 $ & $ 0.13 $ &  HERMES \\
        $ 2458469.3278 $ & $ -28.79 $ & $ -0.30 $ &   $ 1.01 $ & $ 0.16 $ &      ... & ... &  HERMES \\
        $ 2458470.3332 $ & $ -28.71 $ & $ -0.14 $ &  $ 10.21 $ & $ 0.09 $ & $ -10.52 $ & $ 0.14 $ &  HERMES \\
        $ 2458471.3291 $ & $ -28.61 $ &  $ 0.01 $ &  $ 23.17 $ & $ 0.08 $ & $ -26.19 $ & $ 0.13 $ &  HERMES \\
        $ 2458473.3114 $ & $ -28.36 $ & $ -0.43 $ &  $ 23.47 $ & $ 0.09 $ & $ -25.48 $ & $ 0.14 $ &  HERMES \\
        $ 2458474.3081 $ & $ -28.23 $ & $ -0.11 $ & $ -10.97 $ & $ 0.09 $ &  $ 14.33 $ & $ 0.13 $ &  HERMES \\
        $ 2458474.3154 $ & $ -28.44 $ & $ -0.57 $ & $ -10.36 $ & $ 0.53 $ &  $ 14.88 $ & $ 1.39 $ &    NRES \\
        $ 2458475.3148 $ & $ -28.12 $ &  $ 0.00 $ & $ -18.55 $ & $ 0.09 $ &  $ 22.91 $ & $ 0.34 $ &  HERMES \\
        $ 2458476.3158 $ & $ -28.18 $ & $ 0.05 $ & $ -18.67 $ & $ 0.53 $ &  $ 23.28 $ & $ 1.30 $ &    NRES \\
        $ 2458478.2921 $ & $ -27.85 $ & $ -0.37 $ & $ -13.50 $ & $ 0.56 $ &  $ 17.07 $ & $ 1.45 $ &    NRES \\
        $ 2458478.2998 $ & $ -27.86 $ & $ -0.16 $ & $ -13.61 $ & $ 0.27 $ &  $ 17.49 $ & $ 0.31 $ &    NRES \\
        $ 2458478.3074 $ & $ -27.87 $ & $ -1.39 $ & $ -12.37 $ & $ 0.36 $ &  $ 18.33 $ & $ 0.32 $ &    NRES \\
        $ 2458493.3020 $ & $ -24.38 $ & $ -0.30 $ &   $ 3.85 $ & $ 0.61 $ &  $ -3.05 $ & $ 1.52 $ &  HARPSN \\
        $ 2458494.3047 $ & $ -24.10 $ & $ -0.30 $ &  $ 13.53 $ & $ 0.61 $ & $ -14.71 $ & $ 1.54 $ &  HARPSN \\
        $ 2458495.3095 $ & $ -23.82 $ & $ -0.30 $ &  $ 28.73 $ & $ 0.55 $ & $ -32.12 $ & $ 1.46 $ &  HARPSN \\
        $ 2458496.3184 $ & $ -23.53 $ & $ -0.30 $ &  $ 44.43 $ & $ 0.55 $ & $ -50.77 $ & $ 1.46 $ &  HARPSN \\
\hline
\end{tabular}
\end{table*}

\subsection{Orbital parameters determination}
\label{orbitParametersSection}
We use the IDL code \texttt{rvfit} to determine the orbital parameters of the binary system (Iglesias-Marzoa 2015). 
This routine uses an adaptive simulated annealing global minimisation method to fit the two components of the system simultaneously with a Keplerian orbit defined by seven parameters: the orbital period ($P$), the time periastron ($T_{p}$), the eccentricity ($e$), the argument of periastron ($\omega$), the RV amplitude of the two components ($K1$ and $K2$), and the systemic radial velocity ($\gamma$). 
From these we can derive lower bounds for the masses and semi-major axes of the binary components as well as the mass ratio $q=(M_1/M_2)$.
We estimate an initial value for the orbital period from the phase-folded RV values.
The excellent phase coverage of the orbit and the wide time span relative to the orbital period between the CAFE observations and the rest of the exposures allowed us to obtain a very precise initial value, close to the final value provided later on by \texttt{rvfit}. 
The priors and final parameters of the best solution (Fig.~\ref{orbitFigure}) to the simultaneous fit of the two components are shown in Table~\ref{fitResults}.

From this analysis, we determine a value for the systemic RV of GJ\,1284 $ \gamma=0.84\pm0.14\,\mathrm{km\,s}^{-1} $.
This value differs from the one used in the literature to determine the membership of this object to a YMG (Sect.~\ref{DescriptionSection}).
Nonetheless, we note that the individual RV measurements derived by various authors all fall within the range of expected RV values for the primary component at different orbital phases.

\begin{figure}
        \resizebox{\hsize}{!}{\includegraphics{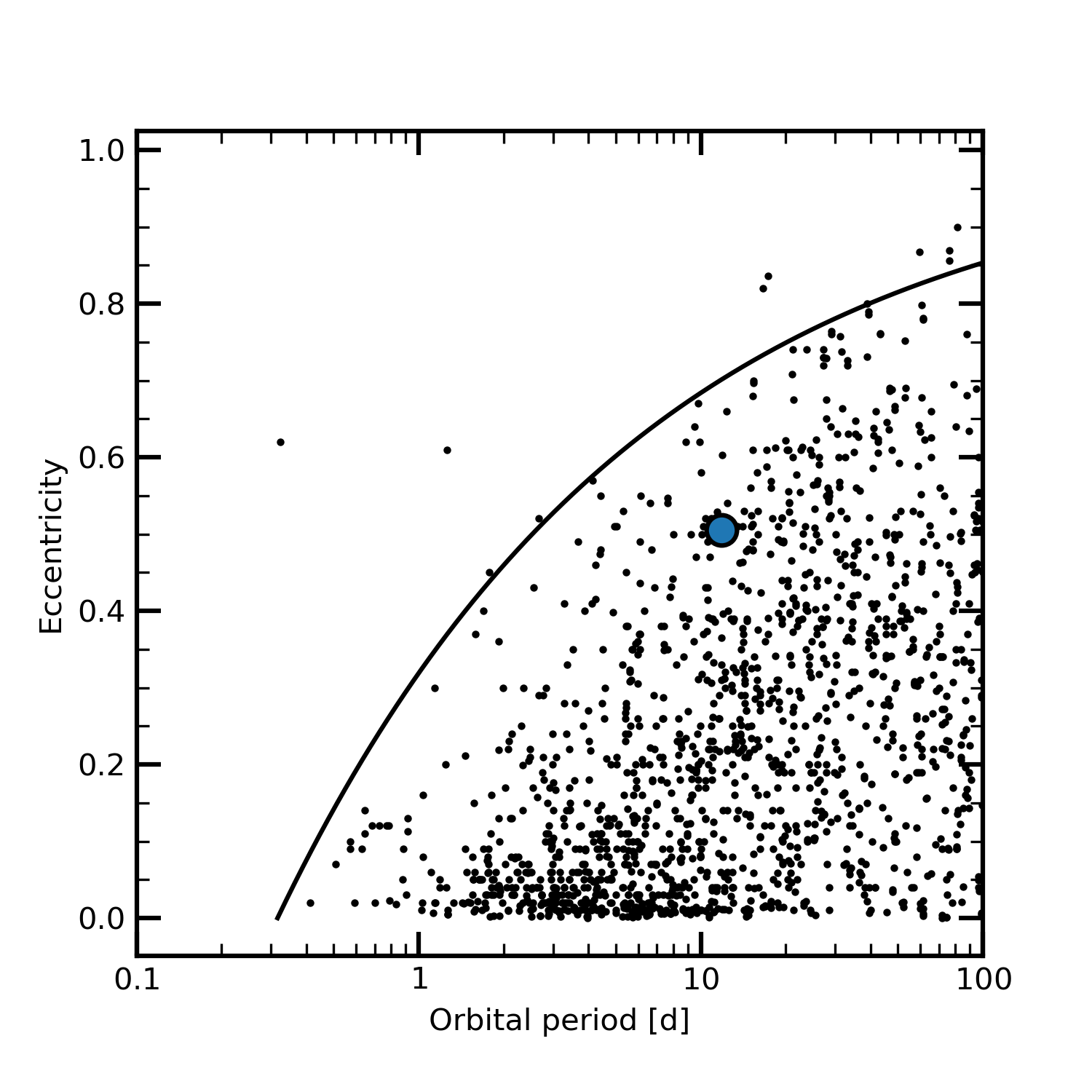}}
        \caption{
                Eccentricity and rotation period of GJ\,1284 (in blue) against binary systems from \citet{Pourbaix2004}. The black solid line represents the upper envelope for this distribution.
        }
        \label{protEccFig}
\end{figure}

We also compare the orbital period and eccentricity of this system against other spectroscopic binaries from \citet{Pourbaix2004} in Fig.~\ref{protEccFig}.
We note that the orbit of GJ\,1284 is highly eccentric for its orbital period, lying close to the upper envelope (solid black line) defined by the binaries of the sample.

\begin{table}
\caption{Parameters of the simultaneous Keplerian fit to both components of GJ\,1284 from \texttt{rvfit}.}
\label{fitResults}
    \centering
    \begin{tabular}{l c c}
    \hline\hline
    \noalign{\smallskip}
         Parameter & Limits & Value \\
    \noalign{\smallskip}
    \hline
    \noalign{\smallskip}
       &  \multicolumn{2}{c}{\textit{Fitted parameters}} \\
    \noalign{\smallskip}
        $ P $ [d]                     & [11.8, 11.9] & $ 11.838033  \pm 0.000016 $ \\
        $ t_{\mathrm{peri}} $ [BJD]   & $-$ & $ 2449997.0000 \pm 0.0012 $ \\
        $ e$                           & [0.0, 0.6] & $ 0.505 \pm 0.005 $ \\
        $ \omega\ [\degr] $            & $ - $ & $ 40.9 \pm 0.3 $ \\
        $ \gamma\ [\mathrm{km\,s^{-1}}] $     & [$-15$, 10] & $ 0.84 \pm 0.14 $ \\
        $K_1\ [\mathrm{km\,s^{-1}}] $        & [20, 50] & $ 31.6 \pm 0.3 $ \\
        $K_2\ [\mathrm{km\,s^{-1}}] $       & [20, 50] &  $ 37.4 \pm 0.3 $ \\
    \noalign{\smallskip}
        & \multicolumn{2}{c}{\textit{Derived quantities}} \\
    \noalign{\smallskip} 
        $ M_1\sin^{3}i\ [\mathrm{M_{\sun}}] $   & & $ 0.141 \pm 0.003 $ \\
        $ M_2\sin^{3}i\ [\mathrm{M_{\sun}}] $   & & $ 0.119      \pm 0.003 $\\
        $ q \equiv (M_1/M_2) $               & & $ 0.845 \pm 0.012 $ \\
        $ a_1\sin{i}\ [\mathrm{R_{\sun}}] $     & & $ 6.38 \pm  0.07 $ \\
        $ a_2\sin{i}\ [\mathrm{R_{\sun}}] $     & & $ 7.56       \pm    0.07 $ \\
     \noalign{\smallskip}
       &  \multicolumn{2}{c}{\textit{Other quantities}} \\
    \noalign{\smallskip}
        $\chi^2$                & & 19.665861 \\
        Time span [day]         & & 1621.6930 \\
        rms$ _1\ [\mathrm{km\,s^{-1}}] $  & & 0.6175     \\
        rms$ _2\ [\mathrm{km\,s^{-1}}] $ & & 0.5919     \\
    \hline& 
    \end{tabular}
\end{table}

\subsection{Rotational period determination}\label{protDetermination}
We determined the rotational period of the three systems from the TESS photometric time series (Sect.~\ref{TESS}).
First, we calculated the generalised Lomb-Scargle (GLS) periodogram \citep{Zechmeister2009} of the photometry and identify the most significant peaks (see central panels of Fig.~\ref{TESSfigure}).
Then, we visually inspected the photometric data folded with the period corresponding to the highest peak.
In  Fig.~\ref{TESSfigure} we can see that GJ\,898 presents a well-defined modulation with a two-peak pattern that is responsible for the peaks at 10.99 and 6.19 days in the periodogram.
However, we find visually that the dispersion in the phase-folded light curve is minimised for a period of $ 11.5 \pm 0.1 $ days, well within the width of the 10.99 days peak,  adopting this value as the rotational period of GJ\,898.
On the contrary, GJ\,1284 and GJ\,897AB present a more complex pattern that can be attributed to the overlapping of the modulation of each component of these two binaries.
Both stars also show intense flare activity, cropped out in Fig.~\ref{TESSfigure} for clarity.

For GJ\,897AB, we find that the shape of the light curve is quite different in the two TESS sectors.
In Sector 2 (in blue in Fig.~\ref{TESSfigure}) we find an irregular modulation with a two-peak pattern showing two main peaks at 5.32 and 2.43 days in the periodogram.
In contrast, the light curve for Sector 29 (in red) shows a clear modulation with a period of 5.02 days.
The periodogram of the combined light curves of the two sectors (in black) shows a peak at 5.00 days.
This suggests that the periodicity at $ \sim $5 days in the light curve is due either to the individual components or to both of them, assuming they have similar rotational period given their similar spectral types and masses.
Thus, we adopted a value of  $ 5.00 \pm 0.02 $\,d as the rotational period of this system.

The light curve of GJ\,1284 also shows a different behaviour for each sector, similarly to GJ\,897.
The light curve of Sector 2 (in blue) presents a more complex structure, while in Sector 29 it is more regular.
The periodograms of both sectors have a main peak at around $ \sim $7.5 days.
We expect a shorter rotation period for the secondary component than for the primary, given its later spectral type   (Sect.~\ref{CMDComparisonSection}), which we do not detect in the TESS light curve as it probably appears diluted due to its lower brightness (Sect.~\ref{massComparisonSection}).
Thus, we determine a rotational period for GJ\,1284A of  $ 7.5 \pm 0.1 $ days.

We followed the same procedure described in this section to calculate the rotation period of the LA members from \citet{Montes2001}.
First, we cross-matched all   120 members with the \text{\it Gaia} eDR3 with a matching radius of 2\arcsec\ accounting for their proper motions to determine the new positions at the \text{\it Gaia} equinox (J2015.5).
We rejected those matches that do not provide a full astrometric solution and radial velocity in \text{\it Gaia}, ending up with a total of 88 stars, for which we derived the UVW velocities and checked that their kinematics are still compatible with a membership to the LA.
Of these 88 stars, 86  have  two-minute-cadence TESS light curves available, but we can only recover a reliable measure of the rotation period for 64 of them, which are listed in Table~\ref{LAProtTable} and plotted in yellow in Fig.~\ref{youthIndicatorsFigure}c.

\section{Discussion} \label{DiscussionSection}
\subsection{Youth assessment}\label{youthSection}
GJ\,1284 has been proposed as a member of $\beta$PMG \citep[$\sim25$\,Myr;][]{Binks2014,Malo2014b} and COL \citep[$\sim40$\,Myr;][]{Torres2008,Bell2015} (Sect.~\ref{DescriptionSection}).
Furthermore, its two wide companions have been kinematically assigned to LA \citep[10--300\,Myr;][]{Lopez-Santiago2009} (Sect.~\ref{WideCompanions}).
The membership of GJ\,1284 to one of these groups, and thus its age, is fundamental so it can be compared  with stellar evolutionary models.

We calculated the UVW galactocentric velocities from the revised systemic RV $ \gamma $ (Sect.~\ref{orbitParametersSection}) and the astrometry from the {\it Gaia} eDR3 using the transformations described in \citet{Johnson1987}.
We also revised the kinematics of GJ\,898 using the astrometry and RV from {\it Gaia} eDR3.
Moreover, we calculated the 3D space motions of GJ\,897AB with the astrometry from \citet{Tokovinin2018} and the RV from \citet{Torres2006} (Sect.~\ref{WideCompanions}).
We show the revised UVW velocities in  Fig.~\ref{UVWplot} and list the values in Table~\ref{stellarParamsTable}.
The kinematics of all three systems agree and are compatible with other members of LA from \citet{Montes2001}.
Moreover, our determination of the systemic velocity of GJ\,1284 place this object outside $ \beta $PMG and COL as defined in the BANYAN-$ \Sigma $ code \citep{Gagne2018}.

We employed several youth indicators to further constrain the kinematic age, including
 the H$ \alpha $ pseudo-equivalent width  $ \mathrm{pEW(H\alpha)} $,
 the near-ultraviolet (NUV) excess emission measured with the colour $ NUV - J $,
 the X-ray luminosity $ L_\mathrm{X} $,
 the stellar rotational period,
 and the pEW of the  Li feature at 6707.8\,nm.

\begin{figure*}
        \centering
        \includegraphics[width=8.2cm]{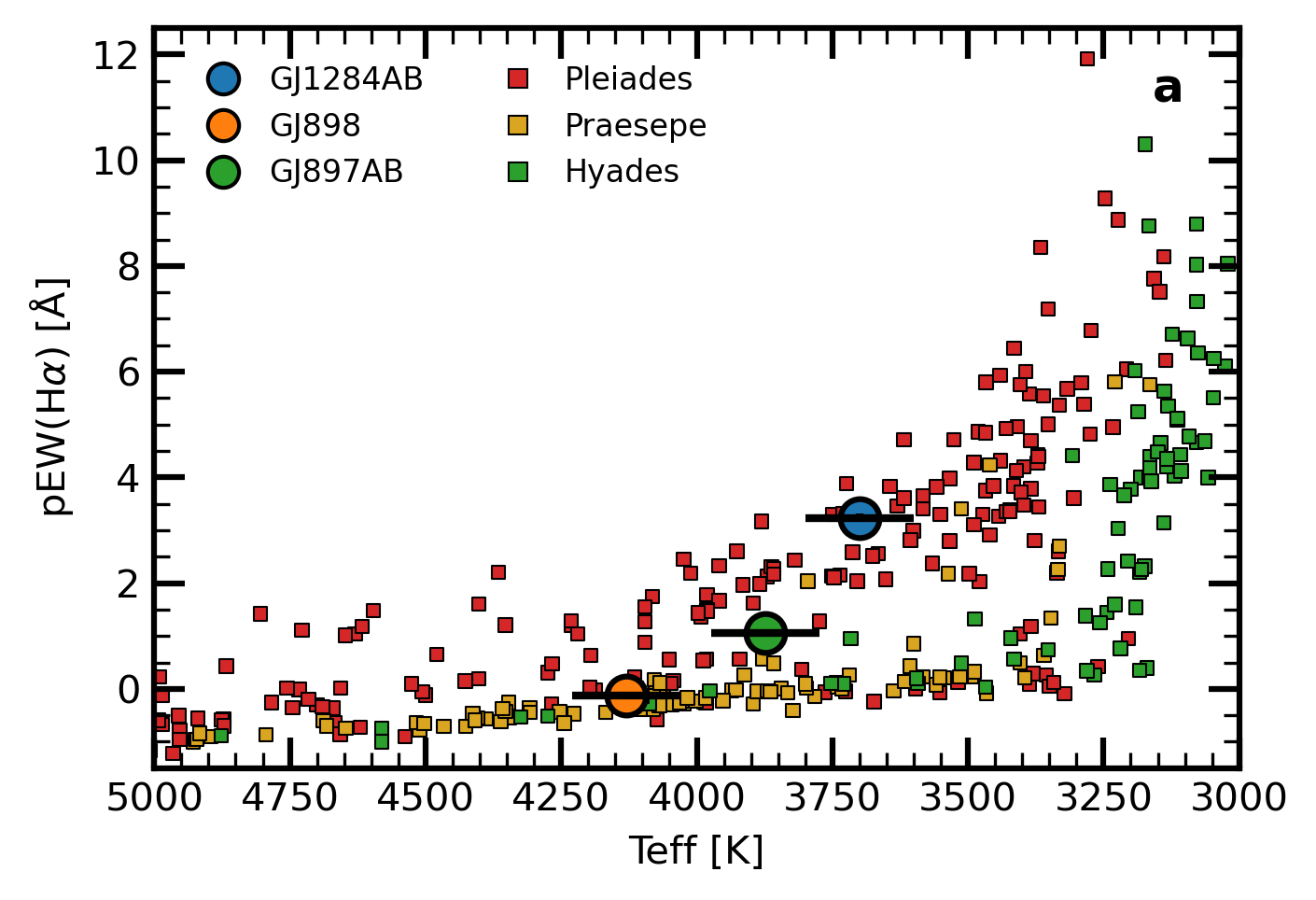}
        \includegraphics[width=8.1cm]{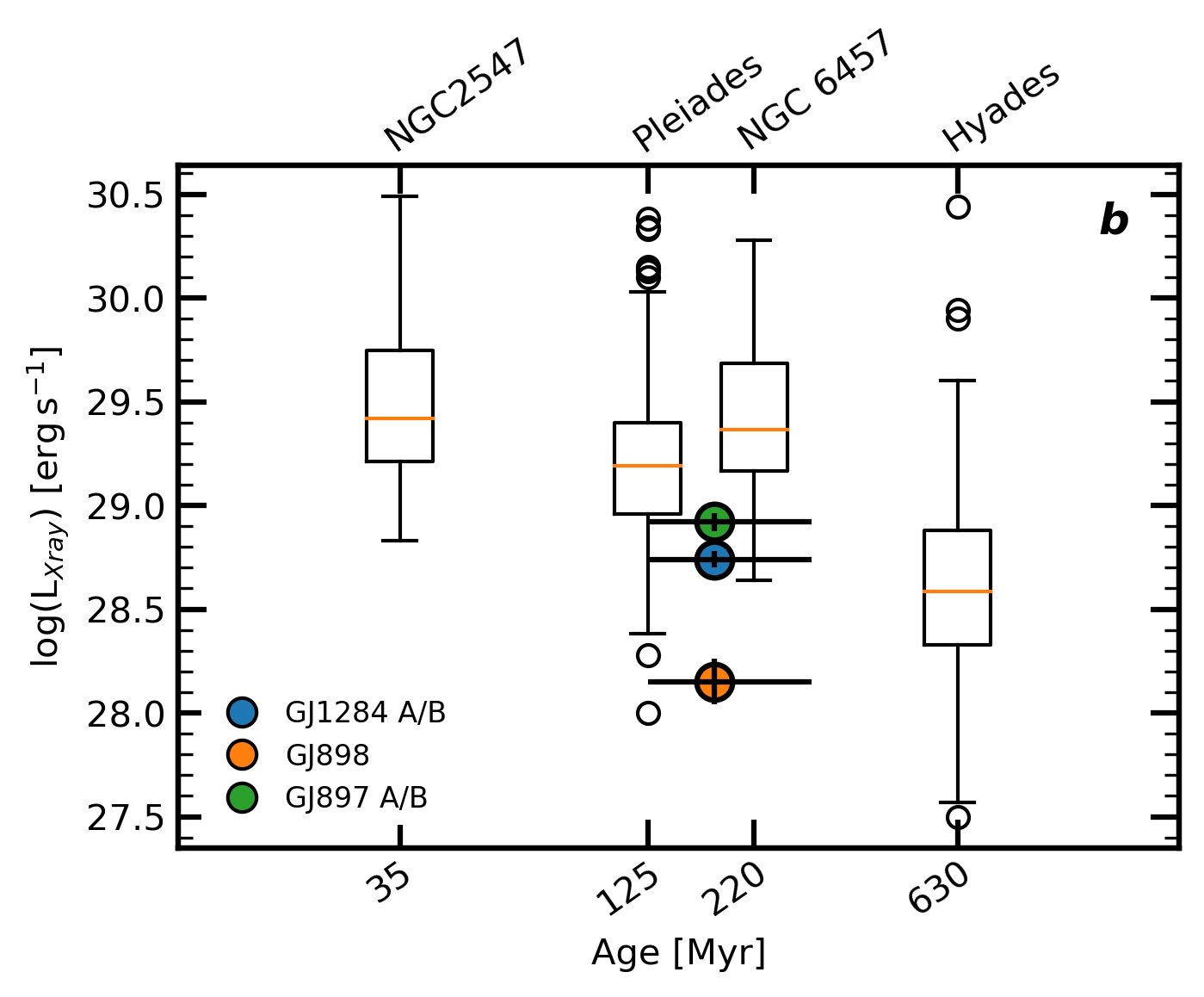}
        \includegraphics[width=8.2cm]{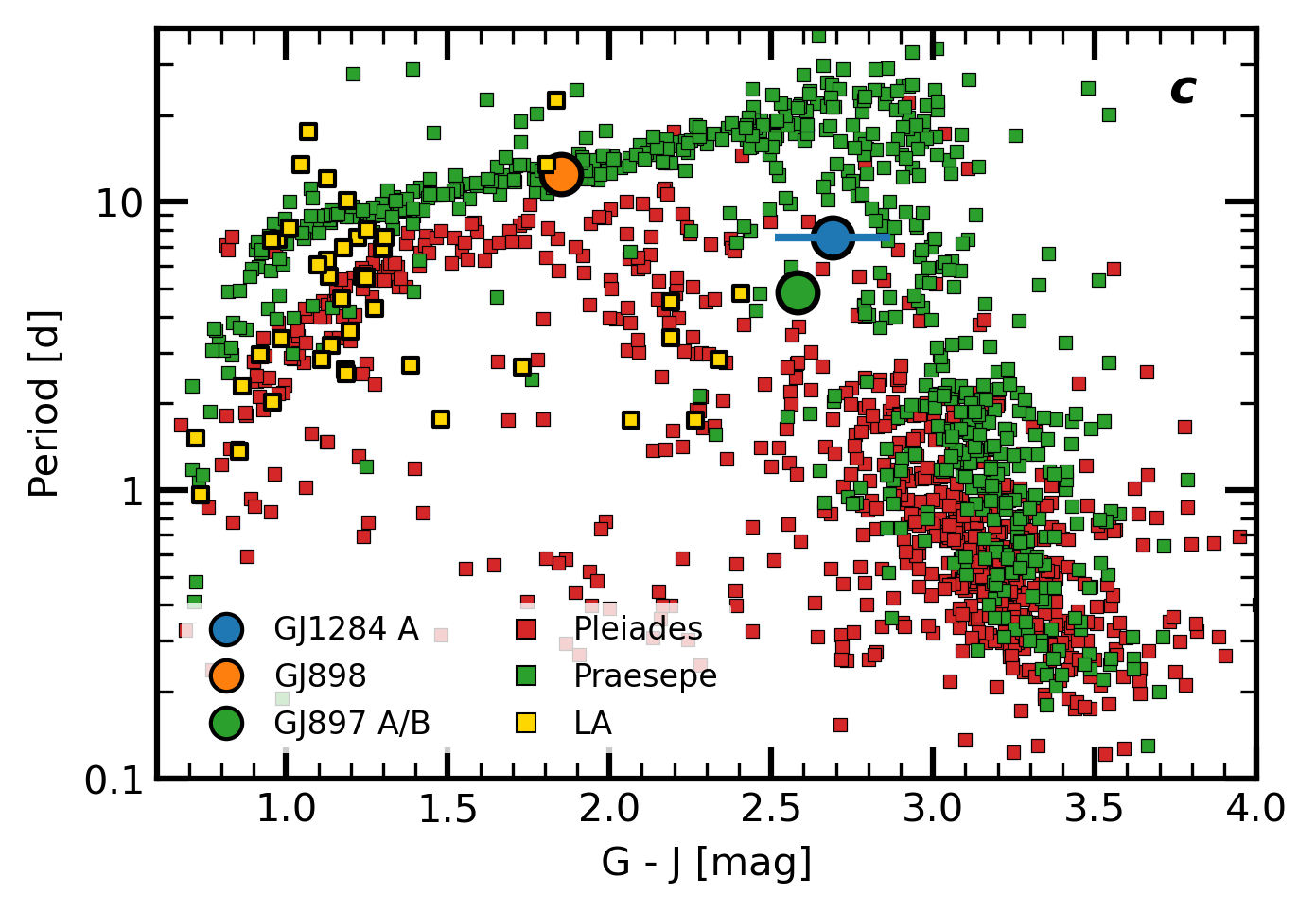}
        \includegraphics[width=8.1cm]{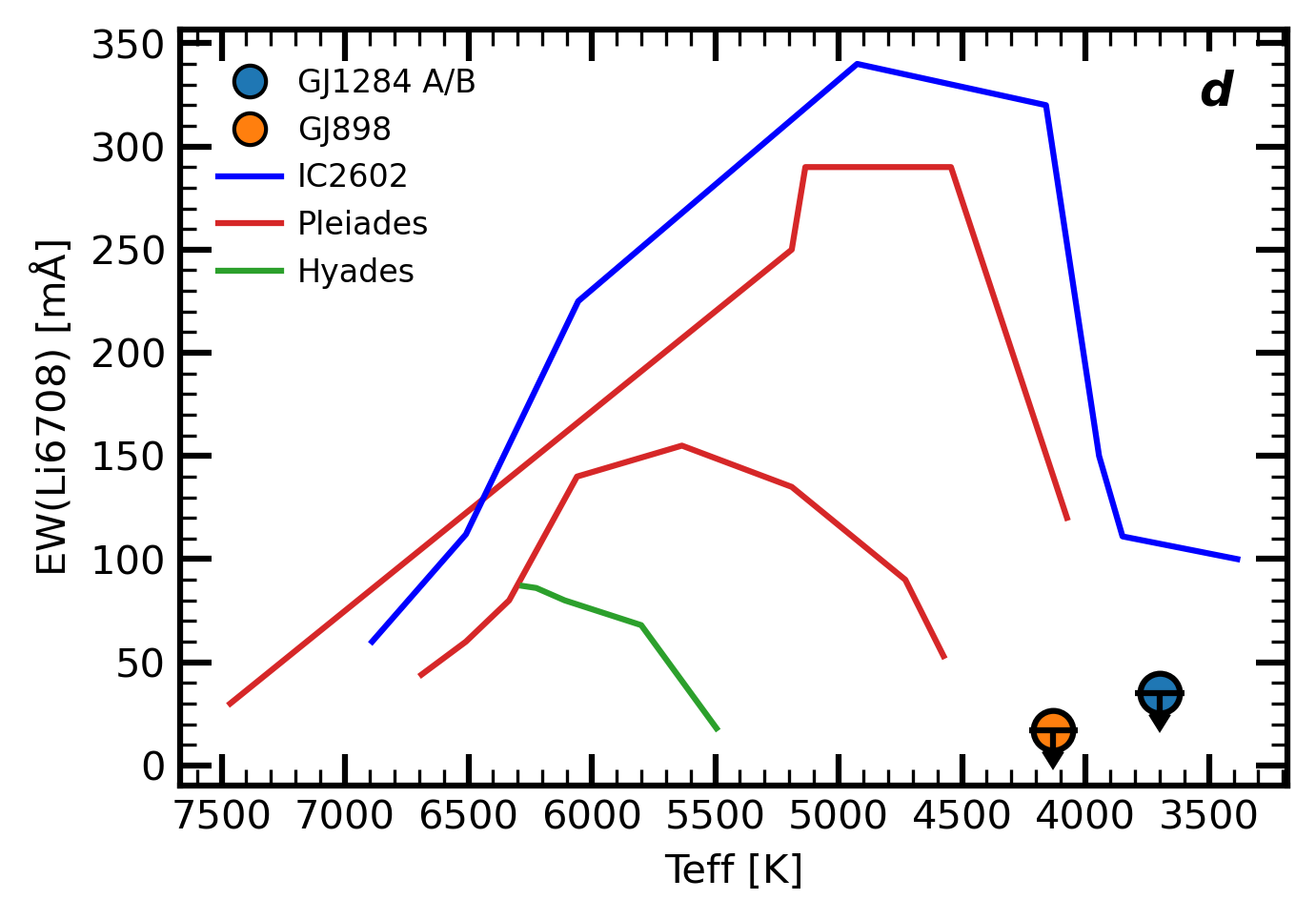}
        \caption{Comparison of the values of four different youth indicators for the three systems analysed in this work (GJ\,1284, GJ\,898, GJ\,897) against young objects from the literature.
        \textit{Panel a}: Pseudo-equivalent width of the H$ \alpha $ line against members of the Pleiades, Praesepe, and Hyades open clusters from \citet{Fang2018};
         \textit{Panel b}: X-ray luminosity of the individual components compared with members from four different open clusters from \cite{Nunez2016} presented in a box-plot;
         \textit{Panel c}: Rotational period derived from the TESS light curves alongside a compilation of members of the Pleiades from \citet{Rebull2016} and of the Hyades from \citet{Douglas2016}, and rotational periods for LA members in \citet{Montes2001} derived in this work (see Sect.~\ref{youthSection});
         \textit{Panel d}: Upper limits of the EW of the Li\,6708\text{\AA} feature plotted alongside the upper envelopes of the IC\,2602 and Hyades open clusters, as well as the upper and lower envelopes of the Pleiades \citep[][and references therein]{GutierrezAlbarran2020}
        }
        \label{youthIndicatorsFigure}
\end{figure*}

In the upper left panel of Fig.~\ref{youthIndicatorsFigure} we show that all three objects present H$ \alpha $ emission compatible with members of the Pleiades \citep[$ \sim $110--130\,Myr;][]{Stauffer1998,Dahm2015}, but also with some stars from the Hyades \citep[$ \sim $650--800\,Myr;][and references therein]{Martin2018,Douglas2019,Lodieu2020b} according to \citet{Fang2018}. We set the age range of the systems between the estimated age of these two open clusters.

From the 2MASS \citep{Skrutskie2006} $ J $ and $ K $ magnitudes and the GALEX \citep{Martin2005} $NUV$ magnitude (Table~\ref{stellarParamsTable}), we inferred colours $ J - K = 0.871, 0.763, 0.854 $\,mag  and $ NUV - J = 11.689, 11.427, 11.395 $\,mag for GJ\,1284, GJ\,898, and GJ\,897, respectively.
\citet{Findeisen2011} identified a correlation between age and NUV excess emission for members of YMGs.
The $ NUV - J  $ colours of our objects indicate an excess emission in the NUV compatible with members of YMGs younger than the Hyades \citep[see Figures~7, 8, and 9 in][]{Findeisen2011}.

We derived a $ \log(L_\mathrm{X}) = 28.15,\ 28.74, \ 28.92\,\mathrm{erg\,s^{-1}}$ for GJ\,898 and the individual components of GJ\,1284AB and GJ\,897AB, respectively, assuming comparable X-ray luminosities for the two components of the binary systems.
These levels of X-ray emission are compatible with objects younger than the Hyades \citep{Nunez2016}, as shown in the upper right panel of Fig.~\ref{youthIndicatorsFigure}.

We compared the rotational periods of GJ\,1284A, GJ\,898, and GJ\,897A against members of the Pleiades and Praesepe open clusters from \citet{Rebull2016} and \citet{Douglas2016,Douglas2019}, respectively  (Fig.~\ref{youthIndicatorsFigure}).
We calculated the $G-J$ colour of GJ\,897A assuming that the two components have similar spectral types.
As for GJ\,1284, the method for calculating the colour of its binary components is described in detail in Sect.~\ref{CMDComparisonSection}.

The lower left panel of Fig.~\ref{youthIndicatorsFigure} shows that the rotational periods of the three stars are longer that the values expected for members of the Pleiades, but shorter than for members of Praesepe \citep[$ \sim $600--750\,Myr;][and references therein]{Douglas2019}.
From the rotational period, we conclude that the age of these three systems is most likely bracketed by the age of the two open clusters.
We note that the rotational period of GJ\,1284 might be affected by the tidal interaction between the two components.
Simulations show that the spin and orbit of binary systems with orbital periods of less than 10 days might synchronise on a timescale as short as 200\,Myr \citep{Fleming2019}.
However, this is not the case for GJ\,1284AB as its orbital period is roughly 1.5 times longer than the rotational period of the primary component.
In any case, the conclusions of the age determination derived from the rotational period of GJ\,898 and GJ\,897A still hold.

We found no presence of the Li6708 line in our GJ\,1284 spectra similarly to \citet{Torres2006} setting an upper limit of $ \mathrm{pEW_{Li6708}}  < 35 $\,m\text{\AA} from our spectra.
We also analysed an archival HARPS spectra for GJ\,898 (programme 085.C-0019(A); PI: Lo Curto)  again finding no sign of the Li6708 line, determining an upper limit for its $ \mathrm{pEW_{Li6708}}  < 17 $\,m\text{\AA}.
This indicates that  their age must be older than IC 2602 \citep[40--50\,Myr][]{Dobbie2010}, and that it  is compatible with members of the Pleiades \citep{GutierrezAlbarran2020} or older, as shown in the lower right panel of Fig.~\ref{youthIndicatorsFigure}.

We  can also put constraints on the age of the GJ\,1284 based on the high eccentricity of its orbit (Table~\ref{fitResults}).
\citet{Meibom2005} determined a relation of the circularisation period with age from binary members of several open clusters.
Although the Hyades cluster appears as an outlier in the original work, a revision of the binaries used in the analysis showed that the cut-off period for this open cluster is in fact 8.5--11.9 days \citep[see][for more details]{Gillen2017}.
The Hyades upper limit is similar to the orbital period of  our system (Table~\ref{fitResults}), thus we infer that  it must  be younger than the age of this open cluster, which agrees with the expected age derived from the other youth indicators.
Moreover, we note again the high eccentricity of the orbit, which can be seen in Fig.~\ref{protEccFig} where GJ\,1284AB is located near the upper envelope, indicating that  it is still far from being circularised and suggesting that  it is well below the age of the Hyades. 
        
We summarise the age range estimated from the lower limit of the younger age and the upper limit from the older age for each activity indicator in Table~\ref{youthRangeTable}.
The upper limit is set by the estimated maximum age of the stellar associations part of LA.
In Fig.~\ref{youthIndicatorsFigure}c we see that the rotation periods of the LA members  from Table \ref{LAProtTable} (in yellow) lie mostly between the sequences defined by the Praesepe and Pleiades stars, which is consistent with the estimated age range of 10--300\,Myr from the literature.
However, we also note that some of them follow the Praesepe sequence or are located slightly above it.
This leads us to believe that the upper limit for the age of this association is not well constrained, although it is indeed young.
Given that the main conclusions from this work are drawn from the fact that the objects have arrived to the main sequence (see Sect.~\ref{CMDComparisonSection}), which is firmly based on the lower limit for the age imposed by the rotational period, the upper limit for the age of LA does not affect them.
Here we adopt a maximum age for the LA similar to those of  Praesepe and Hyades.
The estimated age ranges for the different indicators are then compatible with an age for the system of 110--800\,Myr.

\begin{table}
        \caption{Summary of the possible age ranges for GJ\,1284 according to the different youth indicators considered in this work.}
        \label{youthRangeTable}
        \centering
        \begin{tabular}{l c}
                \hline\hline
                \noalign{\smallskip}
                Youth indicator & Age [Myr]\\
                \noalign{\smallskip}
                \hline
                \noalign{\smallskip}
                Kinematics & 10--800\tablefootmark{a}\\
                H$\alpha$ emission & < 800 \\
                NUV excess emission & < 800 \\
                X-ray luminosity & < 800 \\
                Rotational period & 110--750  \\
                Li 6708\text{\AA} & > 50  \\
                \noalign{\smallskip}
                \hline
                \noalign{\smallskip}
                Final & 110--800 \\
                \hline
        \end{tabular}
        \tablefoot{
                \tablefoottext{a}{
                        The upper limit for the age of the LA members is derived from a comparison of the rotation periods of its members with open clusters. See Sect.~\ref{youthSection} for more details.
                }
        }
\end{table}

\begin{figure*}
        \centering
        \includegraphics[width=17cm]{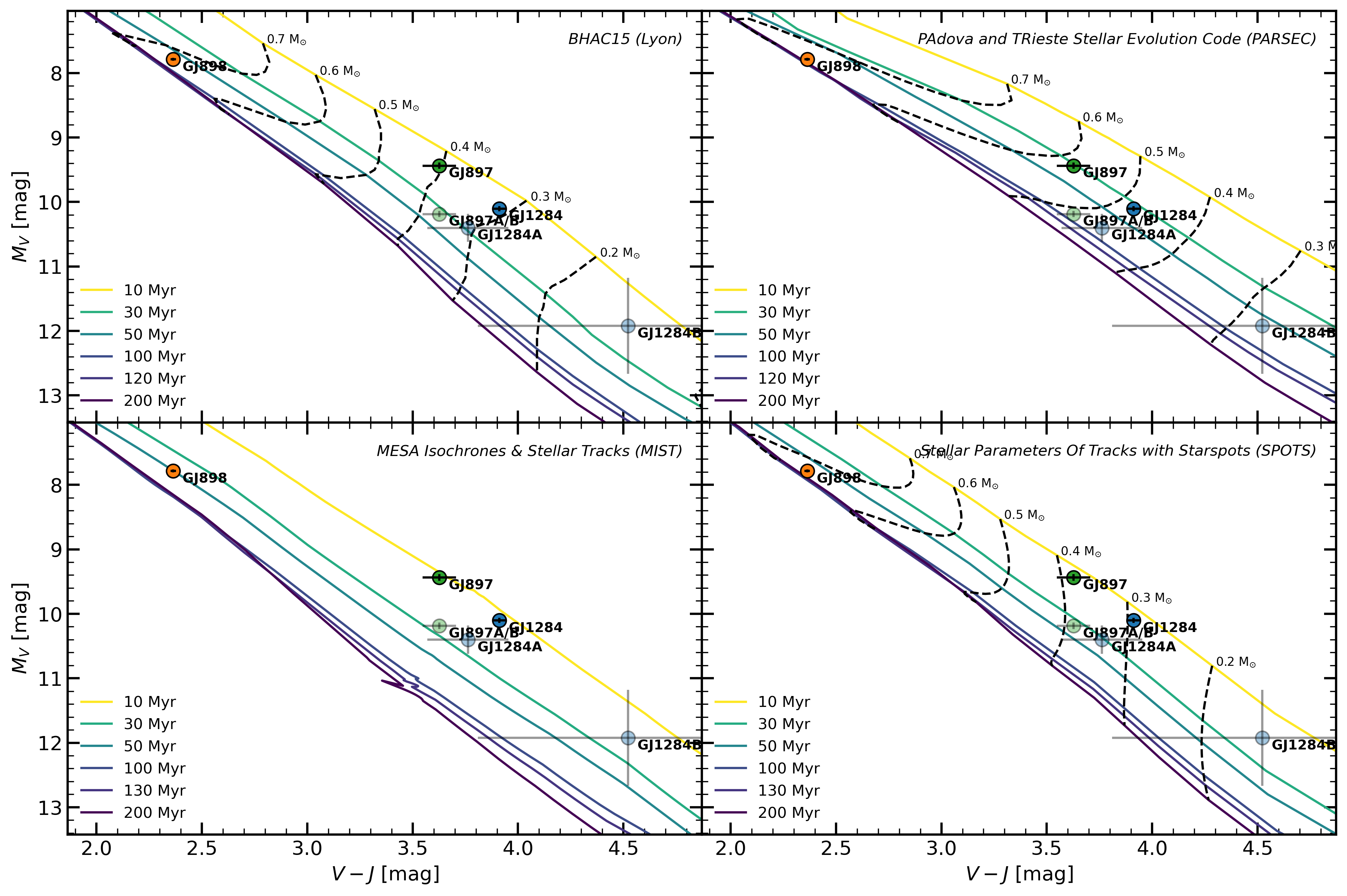}
        \caption{Comparison of the measured magnitudes for GJ\,1284, GJ\,898, and GJ\,897 and the calculated magnitude for the individual components of the two binary systems against a set of isochrones from four different PMS evolutionary models:
        BHAC15 (top left), PARSEC (top right), MIST (bottom left), and SPOTS (bottom right). 
        We also include the isomasses (dashed lines) for each   of the models for initial mass values ranging from 0.7 to 0.2 $M_{\sun}$}
        \label{cmdComparison}
\end{figure*}

\subsection{Gravitational bond of the quintuple system} \label{BondDiscussionSection}
Throughout Sect.~\ref{youthSection} and in the following Sect.~\ref{CMDComparisonSection} we work under the assumption that GJ\,1284 and its two wide companions, GJ\,898 and GJ\,897AB, have a common origin based on their kinematics and thus are roughly the same age.
Even though the youth indicators and kinematics support the membership of these three systems to the LA, given the wide age range of this YMG we cannot confirm they have the same age.
Moreover, the wide physical separation of 1.66\,pc between them might indicate that they are not physically bound.

To explore the different scenarios, we first determine the probability of finding two members of the LA with physical separations and velocities as similar as GJ\,1284 and GJ\,898 based on the galactocentric positions and velocities of the members from \citet{Montes2001}.
We estimate that the probability that two of them have UVW velocities as similar as our systems is $ \sim $ 10--15\%, depending on whether we select all the LA members or only the denser region in the UVW velocity space.
Furthermore, the probability that two LA members are closer than 1.66\,pc is lower than $ 0.2\% $, given that members of YMGs are not grouped as in open clusters, but instead are dispersed over a much larger volume.
Thus, the combined probability of finding two systems at these separations in the position and velocity space is  only $ 0.05\% $ or less.

We also calculate the disruption times of our system to test the hypothesis that they are physically bound using the expressions from \citet{Weinberg1987} and \citet{Binney1987}.
For this calculation we adopt a mean mass for the eventual perturber of 1\,$ M_{\sun} $, a system density of 0.078--0.084\,$ \mathrm{pc}^{-3} $  \citep{Henry2018} and a mean velocity dispersion for the perturbers of 59.8\,$ \mathrm{km\,s^{-1}} $ \citep{ZapateroOsorio2007} for the solar neighbourhood, together with the masses for GJ\,898 ($ \sim 0.66\,M_{\sun} $) and GJ\,1284AB ($ \sim 0.77\,M_{\sun} $, see Sect.~\ref{massComparisonSection}) and their separation (1.66\,pc). 
With these parameters we derive a disruption time $ t_\mathrm{disr} \simeq  $100--200\,Myr, which is comparable to the age estimated from the youth indicators (Sect.~\ref{youthSection}) suggesting that the system might still be physically bound.
However, the difference in the W velocities of GJ\,1284 and GJ\,898, an order of magnitude higher than the escape velocity of the latter of $ \sim 100\,\mathrm{m\,s^{-1}} $, indicates otherwise.
Nonetheless, if we trace back the motion of these two systems using their UVW velocities, we find that they reached their minimum separation (0.6\,pc) just $ 1.3 \pm 0.5 $\,Myr ago.
Using this distance instead, we infer a disruption time $ t_\mathrm{disr} \simeq $400--700\,Myr, longer than the time we calculated previously, but still comparable to the age of the system.
In summary,  although currently the system does not seem to be physically bound, we believe that  it is very unlikely to find such relatively close systems in the LA by chance alignment.
Additionally, we cannot rule out that the system was physically bound until 1--2\,Myr ago when it could have been disrupted by an impact from a perturber.

\subsection{Model comparison}
\subsubsection{Colour-magnitude diagrams} \label{CMDComparisonSection}
In this section we compare the absolute magnitudes and colours of the three systems with four different PMS evolutionary models:
 Lyon models \citep[BHAC15;][]{Baraffe2015}; MESA Isochrones \& Stellar Tracks \citep[MIST;][]{Dotter2016,Choi2016}; PAdova and TRieste Stellar Evolution Code \citep[PARSEC;][]{Bressan2012,Chen2014}; and Stellar Parameters of Tracks with Starspots \citep[SPOTS;][]{Somers2020}. A brief description of the characteristics of each of these models can be found in \citet{David2019}.
In  Fig.~\ref{cmdComparison} we plot the absolute $ V $ magnitudes and the $ V - J $ colour of each system, and of their individual components together with the isochrones from each theoretical model.

To separate the integrated photometry of GJ\,1284 into the corresponding magnitudes for the individual components we use the integrated $ Gaia\ G_\mathrm{BP} $ magnitude and the flux ratio derived by \texttt{TODMOR} from the HARPS-N spectra.
The wavelength range of the HARPS-N spectra closely resembles the transmission curve of the $ G_\mathrm{BP} $ filter\footnote{\url{https://www.cosmos.esa.int/web/gaia/iow_20180316}}, hence justifying this approximation.
The absolute $ G_\mathrm{BP} $ is calculated from the parallax of this object (Table~\ref{stellarParamsTable}).
Using these values, we derive absolute magnitudes in the $ G_\mathrm{BP} $ filter of $ 10.73 \pm 0.09 $ and $ 11.55 \pm 0.13$\,\mbox{mag} for the primary and secondary components, respectively.
Additionally, we use the relations for old field stars from \citet{Pecaut2013}\footnote{See updated version including Sloan and {\it Gaia} filters in \url{http://www.pas.rochester.edu/~emamajek/EEM_dwarf_UBVIJHK_colours_Teff.txt}} assuming that the system is close to the main sequence based on its age (Sect.~\ref{youthSection}).
We base this approximation on the fact that, for solar-type stars and  early- to mid-type M dwarfs, the isochrones of the evolutionary models for ages older than $ \sim $120\,Myr, as our objects appear to be, squeeze together (Fig.~\ref{cmdComparison}).
Moreover, a comparison with empirical masses and luminosities of members of the NGC\,1647 \citep[$ \sim $150\,Myr;][]{Dias2002}, Pleiades, and Praesepe open clusters with similar spectral types  (Sect.~\ref{massComparisonSection}) show that the mass-luminosity sequences overlap at this range of ages.
According to the $ G_\mathrm{BP} $ absolute magnitude, we infer a spectral type of M2--M2.5 and M3--M3.5 respectively for the primary and secondary components of GJ\,1284 using these relations.
We estimate the difference in magnitudes between the two components for the different filters from these same relations and decompose the other integrated photometry into the individual magnitudes.
We take into account the uncertainties in the spectral type estimates to calculate the photometric error bars.
For GJ\,897AB, the two components of the binary system have similar spectral types (Sect.~\ref{WideCompanions}), hence we assume identical fluxes. 

Figure~\ref{cmdComparison} shows that the BHAC15, MIST, and SPOTS models predict a younger isochronal age for GJ\,1284A/B and GJ\,897A/B ($ \sim $30--50\,Myr) than for GJ\,898 ($ \sim $100\,Myr).
This effect was already noticed by \citet{Herczeg2015} using the BHAC15 models, which predicted younger ages for very low-mass stars than for solar-type stars in other YMGs.
In Table~\ref{isochroneAgeTable} we also list the isochronal age calculated minimising the combined distance of the three systems to each isochrone assuming they are coeval.
To account for the uncertainties in the magnitude and colour of the individual components we performed a Monte Carlo simulation with 5\,000 realisations to estimate the error in the age determination.
We see that these ages are younger than those derived from the youth indicators in Sect.~\ref{youthSection}, in particular from the rotational period.
The determination of isochronal ages younger than expected for members of other young associations  was also noted in other works \citep[e.g.][]{David2016,Gillen2017}.   
On the other hand, PARSEC models suggest a common age of $ \sim $40--140\,Myr for all of them, consistent with the youth indicators.
The version v1.2S of the PARSEC models considered in this work includes an ad hoc correction to reconcile the models with the observational data, particularly for very low-mass stars \citep{Chen2014}.
This might explain the differences between PARSEC and the other three sets of models.

\begin{table}
        \caption{Summary of the isochronal ages for each model derived for the three systems analysed in this work, assuming coevality.}
        \label{isochroneAgeTable}
        \centering
        \begin{tabular}{l c c c c}
                \hline\hline
                \noalign{\smallskip}
                        &  BHAC15  &  PARSEC  &  MIST  &  SPOTS\\
                \noalign{\smallskip}
                \hline
                \noalign{\smallskip}
                Age [Myr] & $ 35^{+25}_{-14} $ &  $ 72^{+67}_{-35} $  &  $ 30^{+25}_{-13} $  &  $ 42^{+48}_{-22} $  \\
                \noalign{\smallskip}
                \hline
        \end{tabular}
\end{table}

\begin{figure}
        \resizebox{\hsize}{!}{\includegraphics{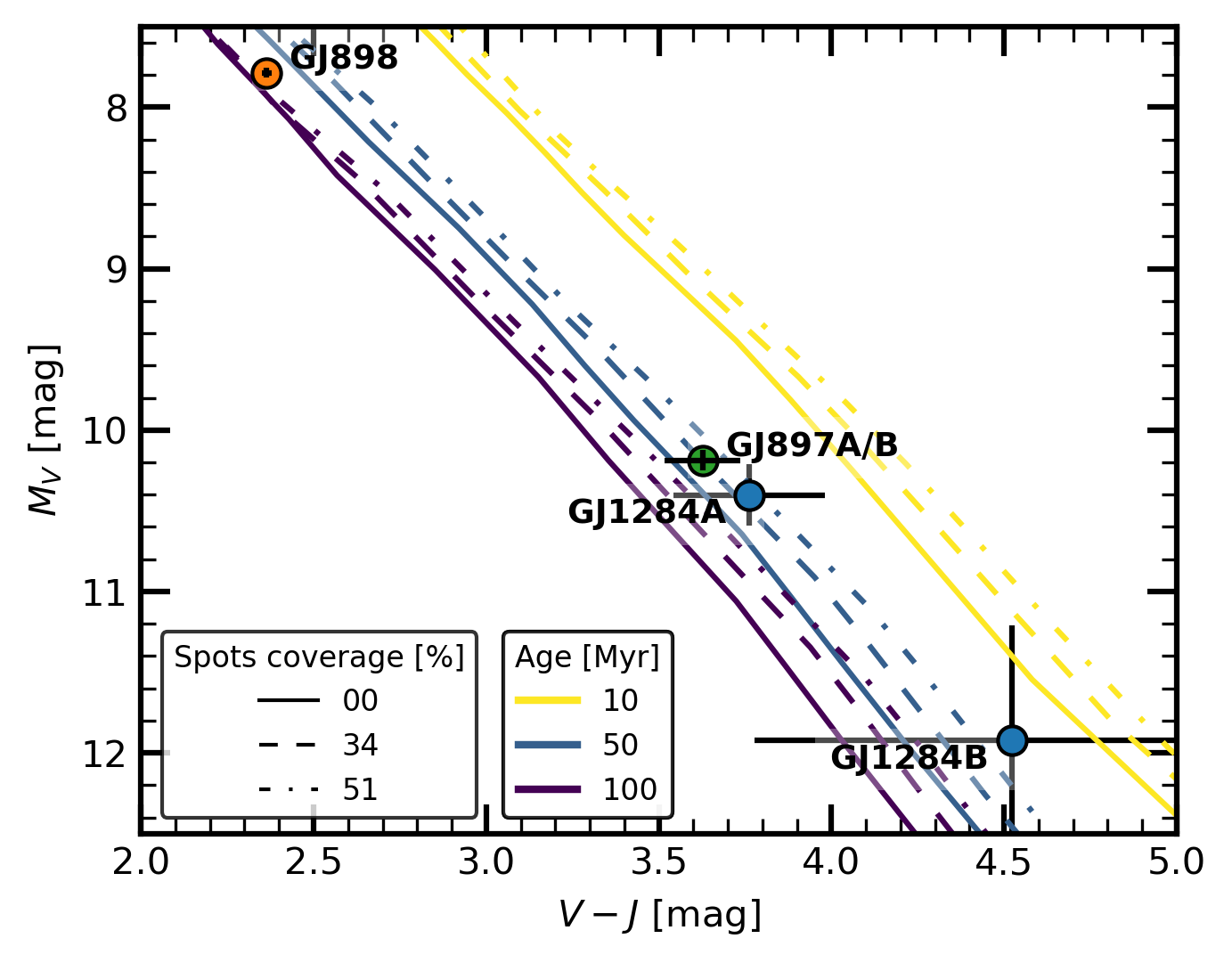}}
        \caption{Isochrones from the SPOTS model for three different ages (10, 50, and 200 Myr) and different spot-filling factors (0, 34, and 51\%). }
        \label{spotsComparison}
\end{figure}

The discrepancy in the age derived by the different models indicates that there might be an additional factor that varies for the three systems.
For example, the SPOTS models include the effects of magnetic fields in the form of stellar spots, while the other models do not \citep{Somers2020}.
\citet{Somers2015} explored in detail the effect of spot-related activity in the stellar evolution and found that low-mass stars with starspots have lower luminosities and temperatures compared to those with no spots.
As a consequence, ages in spotted stars can be underestimated by a factor of 2--10.
However, the spot-filling factor cannot be directly determined from our observations.
In  Fig.~\ref{spotsComparison} we show isochrones derived from the SPOTS models for different spot-filling factors.
We can see  for the typical maximum value of 50\% in PMS low-mass stars (\citealt{Somers2015} and reference therein) that the isochronal ages of the three systems are in better agreement with the lower limit for the age $ \sim $100\,Myr derived from the youth indicators and the values derived by the PARSEC models.

\subsubsection{Mass-luminosity relations} \label{massComparisonSection}

\begin{figure*}
        \centering
        \includegraphics[width=17cm]{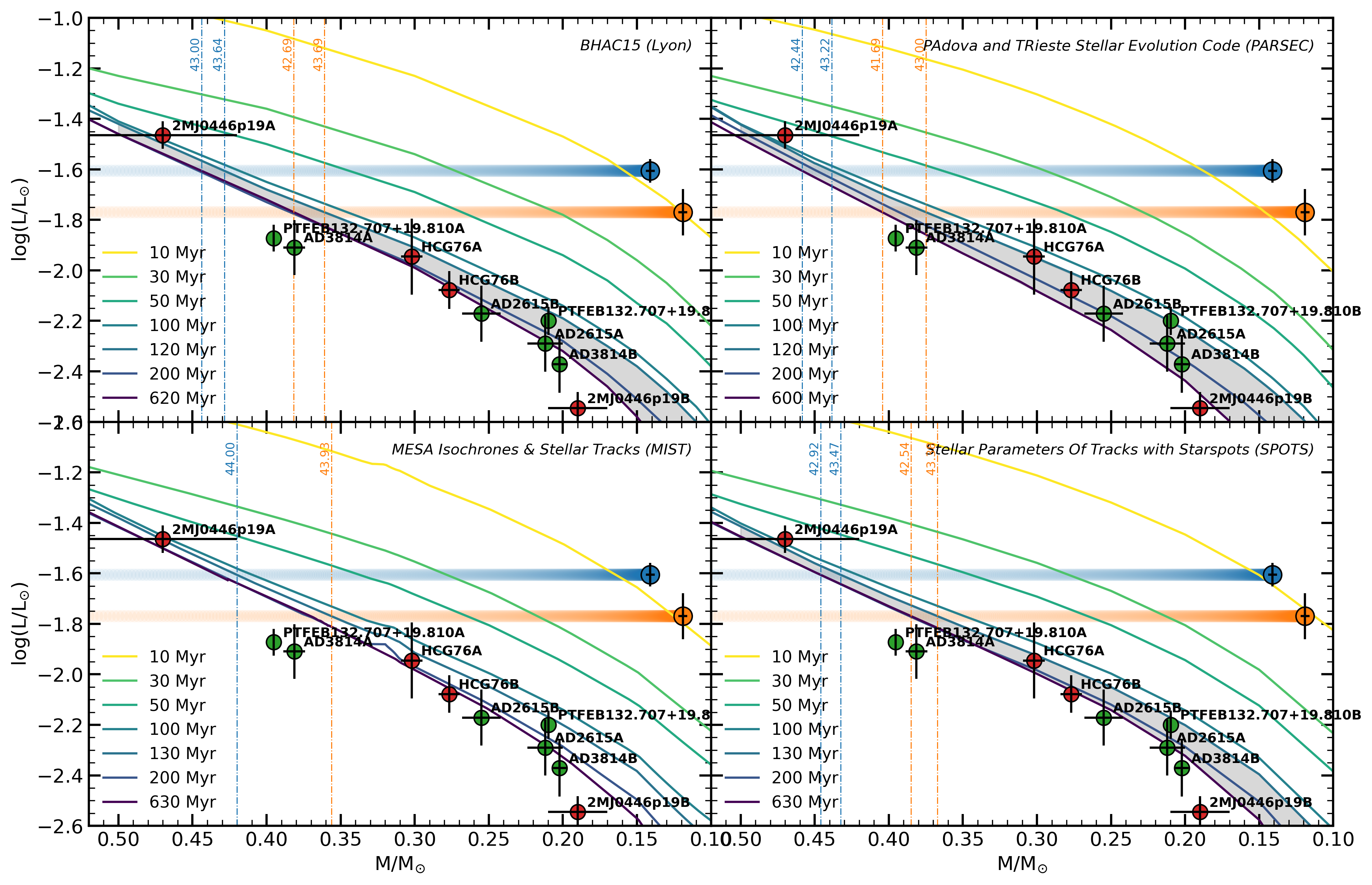}
        \caption{
                Minimum masses and luminosities of GJ\,1284 A (blue) and B (orange) against the mass-luminosity relations derived from the BHAC15 (top left), PARSEC (top right), MIST (bottom left), and SPOTS (bottom right) models.
                The blue and orange horizontal bars represent the expected masses at different inclination angles for the primary and the secondary components, respectively.
                The dashed vertical lines indicate the masses derived from the mass-luminosity relations for the minimum and maximum ages derived from the youth indicators (shaded in grey).
                For comparison, we overplot the dynamical masses of the EB members of NGC\,1647, the Pleiades (red) and Praesepe (green) open clusters from \citet{Lodieu2020a}.
        }
        \label{massComparisonFigure}
\end{figure*}

In this section we compare the minimum masses of GJ\,1284A and B (Sect.~\ref{orbitParametersSection}) and other dynamical mass determinations with the predictions from the PMS models in a mass-luminosity diagram (Fig.~\ref{massComparisonFigure}).
We use the bolometric corrections from \citet{Pecaut2013} for the spectral types of the individual components and 
the $ G_\mathrm{BP}$ magnitudes of GJ\,1284A and B (Sect.~\ref{CMDComparisonSection}) to derive a bolometric luminosity of  $ \log( L_\mathrm{bol } /  L_{\sun})= -1.61 \pm 0.05 $ and $ -1.77 \pm 0.09 $ for the primary and the secondary components, respectively.
These values for the luminosities and the minimum masses are compatible with the theoretical mass-luminosity relations.

Given the bolometric luminosity of the individual components of GJ\,1284, we can estimate their masses using the mass-luminosity relations from the models assuming an age range for the system of 110--800\,Myr (Sect.~\ref{youthSection}) obtaining a mean mass of $ 0.432 $--$ 0.462\,M_{\sun} $ and $ 0.365 $--$ 0.390\,M_{\sun} $ for the primary and secondary, respectively.
The intervals of masses for each model (Table~\ref{isochroneMassTable}) take into account the estimated range of ages.
We compare these results with dynamical masses measured for EBs with similar ages and spectral types (Fig.~\ref{massComparisonFigure}) from a compilation by \citet{Lodieu2020a}, namely with members of the Pleiades (in red) from \citet{David2016} and \citet{Hebb2006} and Praesepe (in green) from \citet{Kraus2017} and \citet{Gillen2017} with spectral types M2--M4 for at least one of their components.
We see that the masses derived for GJ\,1284A and B from their luminosity and age are consistent with the dynamical masses for these EBs, and in particular they seem to be delimited between the masses of 2MASSJ0446+19A \citep[M2.5V; $ 0.47 \pm 0.05\,\mathrm{M_{\sun}}$;][]{Hebb2006} and PTFEB132.707+19.810 \citep[M3.5V; $ 0.3953 \pm 0.0020\,\mathrm{M_{\sun}}$;][]{Kraus2017}, whose spectral types are also compatible with the GJ\,1284 components (Sect.~\ref{CMDComparisonSection}).
Assuming that the masses of each component derived from the photometry are correct, they imply an inclination angle for the orbit of the binary system of $ 43.6 $--$ 42.8\degr $ or $ 43.5 $--$ 42.3\degr $ according to the minimum mass of primary and the secondary, respectively.

\begin{table}
        \caption{Summary of maximum and minimum masses (in $ \mathrm{M_{\sun}} $) of the two binary components derived from the isochrones of the different models.}
        \label{isochroneMassTable}
        \centering
        \begin{tabular}{l c c c c}
                \hline\hline
                \noalign{\smallskip}
                Model & \multicolumn{2}{c}{Primary} & \multicolumn{2}{c}{Secondary}\\
                \noalign{\smallskip}
                & Min & Max & Min & Max\\
                \noalign{\smallskip}
                \hline
                \noalign{\smallskip}
                BHAC15 &   0.427 &   0.452 &   0.361 &   0.382 \\
                PARSEC &   0.444 &   0.479 &   0.375 &   0.404 \\
                MIST   &   0.422 &     0.440 &   0.356 &     0.372  \\
                SPOTS  &   0.435 &   0.456 &   0.367 &   0.385 \\
                \hline
        \end{tabular}
\end{table}

We can also see, with the exception of PARSEC, that the masses derived from the mass-luminosity relations for each model are higher than the values inferred from the CMD for the same model.
Moreover, the masses derived from the CMD are lower than the dynamical masses from the literature.
The colour and magnitude corrections needed to reconcile the masses derived from the mass-luminosity relations and the CMDs are listed in Table~\ref{magColourDiffsTable}.
This suggests that the mass-luminosity relations seem to be more reliable to derive masses from the photometry than CMDs.
One of the possible explanations for this could be that, although the evolutionary models may correctly predict the values of the luminosities, they appear to underestimate the effective temperatures.
This was already noted in other works \citep{David2016,Gillen2017}, where models predicted effective temperatures 200--300\,K hotter for early to mid M dwarfs.
Alternatively, model atmospheres and spectral synthesis  for these objects may predict magnitudes and colours for these very cool objects that are not fully correct.
Furthermore, this also indicates that, unless the age of the system is younger than the value derived from the youth indicators in this work, only the PARSEC models seem to predict masses consistent with those derived from mass-luminosity relations and CMDs and also compatible with empirical masses of EBs in open clusters older than 100\,Myr.

\begin{table}
        \caption{Colour and magnitudes differences needed to reconcile the mass estimates derived from the CMD and the mass-luminosity relations for each model.}
        \label{magColourDiffsTable}
        \centering
        \begin{tabular}{l c c c c}
                \hline\hline
                \noalign{\smallskip}
                Model & \multicolumn{2}{c}{Primary} & \multicolumn{2}{c}{Secondary}\\
                \noalign{\smallskip}
                & $ \Delta M_V $ & $ \Delta (V-J) $ & $ \Delta M_V $ & $ \Delta (V-J) $ \\
                \noalign{\smallskip}
                & [mag] & [mag] & [mag] & [mag]\\
                \noalign{\smallskip}
                \hline
                \noalign{\smallskip}
                BHAC15 &   0.13 &   0.48 &   1.03 &   1.02 \\
                PARSEC &  -0.08 &   0.19 &   0.64 &   0.62 \\
                MIST   &   0.27 &   0.67 &   1.16  &   1.21 \\
                SPOTS  &   0.05 &   0.41 &  0.89  &  0.90 \\
                \hline
        \end{tabular}
        \tablefoot{
        The uncertainties associated with these colour and magnitude differences are dominated by the error on the magnitudes (0.19, 0.71\,mag ) and the colours (0.21, 0.74\,mag) of the primary and secondary components, respectively.
        }
\end{table}

\section{Conclusions}
In this work we analyse a young SB2 system, GJ\,1284, alongside its two wide companion, GJ\,898 and GJ\,897AB, and compare the results of this analysis with several PMS evolutionary models.
The main results and conclusions from this work are as follows:
\begin{itemize}
        \item{
                We characterised the orbit of GJ\,1284, a young SB2 system, using RVs derived using high-resolution spectra from HERMES, HARPS-N, CAFE, and NRES.
                We find that this system is composed of two stars of  similar mass ($ q = 0.845 \pm 0.012 $)  with estimated spectral types of M2--M2.5 for the primary component and M3--M3.5 for the secondary in an eccentric orbit ($ e = 0.505 \pm 0.005 $) with a period of $ P = 11.838033 \pm 0.000016 $\,d.
                We also derived a minimum mass of $ 0.141 \pm 0.003,\ 0.119 \pm 0.003 \,\mathrm{M_\sun} $ for the primary and secondary components, respectively.
        }
        \item{
                We re-analysed the kinematics of GJ\,1284AB using the revised systemic velocity of $ 0.84 \pm 0.14\,\mathrm{km\,s^{-1}} $, which suggests a membership in the LA, consistent GJ\,898 and GJ\,897AB, in the same YMG.
                Based on several youth indicators, we adopted an age range for the three systems of 110--800\,Myr, assuming they have a common origin.
        }
        \item{
                We find that the probability of finding a system such as the one made up of  GJ\,1284 and its two relatively close co-moving companions in the LA by chance alignment is very low, suggesting they might have a common origin.
                Moreover, although currently they do not seem to be physically bound, we cannot rule out that they were in the past before being disrupted just 1--2\,Myr ago.
        }
        \item{
                We find that the isochronal ages derived from the comparison of the photometry with four PMS evolutionary models (BHAC15, PARSEC, MIST, and SPOTS) derived from the photometry of these three systems are younger than those estimated from the youth indicators, except for the PARSEC models.
                Moreover, the isochrones of the models, again excluding PARSEC, suggest a different age for the early M dwarf binaries than for the solar-type star.
        }
        \item{
                Finally, we see that the masses for GJ\,1284 A and B from the mass-luminosity relations for the different evolutionary models, except PARSEC, are higher than the values derived from the photometry alone, but are consistent with the dynamical masses of EBs with similar spectral type and age from the literature.
                This suggests that the determination of masses using mass-luminosity relations from evolutionary models may be more reliable than a comparison with the photometry in CMDs.
                Alternatively, the system might be younger than  we estimated from the youth indicators.
        }
\end{itemize}

\begin{acknowledgements}
We thank the anonymous referee for his/her comments that helped to improve this work.

CCG is supported by the Ministerio de Economía y Competitividad del Gobierno de España (MINECO) under project SEV-2015-0548-16-3.

We acknowledge financial support from the Agencia Estatal de Investigación of the Ministerio de Ciencia, Innovación y Universidades through projects 
PID2019-109522GB-C5[3, 4] and AEI/10.13039/501100011033.

Based on observations made with the Mercator Telescope, operated on the island of La Palma by the Flemish Community, at the Spanish Observatorio del Roque de los Muchachos of the Instituto de Astrofísica de Canarias. 

Based on observations obtained with the HERMES spectrograph, which is supported by the Research Foundation - Flanders (FWO), Belgium, the Research Council of KU Leuven, Belgium, the Fonds National de la Recherche Scientifique (F.R.S.-FNRS), Belgium, the Royal Observatory of Belgium, the Observatoire de Genève, Switzerland and the Thüringer Landessternwarte Tautenburg, Germany. 

This work has made use of data from the European Space Agency (ESA) mission {\it Gaia} (\url{https://www.cosmos.esa.int/gaia}), processed by the {\it Gaia} Data Processing and Analysis Consortium (DPAC, \url{https://www.cosmos.esa.int/web/gaia/dpac/consortium}).
Funding for the DPAC has been provided by national institutions, in particular the institutions participating in the {\it Gaia} Multilateral Agreement.

This work makes use of observations from the NRES spectrograph on the LCOGT 1m telescope at the SAAO node of the Las Cumbres Observatory global telescope network.

Based on observations made with the Italian Telescopio Nazionale Galileo (TNG) operated on the island of La Palma by the Fundación Galileo Galilei of the INAF (Istituto Nazionale di Astrofisica) at the Spanish Observatorio del Roque de los Muchachos of the Instituto de Astrofisica de Canarias.
 
 Based on observations collected at the Centro Astronómico Hispano-Alemán (CAHA) at Calar Alto, operated jointly by Junta de Andalucía and Consejo Superior de Investigaciones Científicas (IAA-CSIC).
 
 This paper includes data collected by the TESS mission. Funding for the TESS mission is provided by the NASA's Science Mission Directorate.
 
 This publication makes use of data products from the Two Micron All Sky Survey, which is a joint project of the University of Massachusetts and the Infrared Processing and Analysis Center/California Institute of Technology, funded by the National Aeronautics and Space Administration and the National Science Foundation.
 
\end{acknowledgements}


\bibliographystyle{aa} 
\bibliography{gj1284_ccardona2021} 

\begin{thebibliography}{112}
\expandafter\ifx\csname natexlab\endcsname\relax\def\natexlab#1{#1}\fi

\bibitem[{Aceituno {et~al.}(2013)Aceituno, {F S{\'{a}}nchez}, Grupp, Lillo,
  Hern{\'{a}}n-Obispo, Benitez, Montoya, Thiele, Pedraz, Barrado, Dreizler, \&
  Bean}]{Aceituno2013}
Aceituno, J., {F S{\'{a}}nchez}, S., Grupp, F., {et~al.} 2013, A{\&}A, 552, A31

\bibitem[{Asiain {et~al.}(1999)Asiain, Figueras, Torra, \& Chen}]{Asiain1999}
Asiain, R., Figueras, F., Torra, J., \& Chen, B. 1999, A{\&}A, 341, 427

\bibitem[{Baglin {et~al.}(2006)Baglin, Auvergne, Boisnard, Lam-Trong, Barge,
  Catala, Deleuil, Michel, \& Weiss}]{Baglin2006}
Baglin, A., Auvergne, M., Boisnard, L., {et~al.} 2006, in 36th COSPAR Sci.
  Assem., 3749

\bibitem[{Baraffe {et~al.}(2015)Baraffe, Homeier, Allard, \&
  Chabrier}]{Baraffe2015}
Baraffe, I., Homeier, D., Allard, F., \& Chabrier, G. 2015, A{\&}A, 577, A42

\bibitem[{{Barrado y Navascu{\'{e}}s} {et~al.}(1999){Barrado y
  Navascu{\'{e}}s}, Stauffer, Song, \& Caillault}]{BarradoyNavascues1999b}
{Barrado y Navascu{\'{e}}s}, D., Stauffer, J.~R., Song, I., \& Caillault, J.-P.
  1999, ApJ, 520, L123

\bibitem[{Bell {et~al.}(2015)Bell, Mamajek, \& Naylor}]{Bell2015}
Bell, C. P.~M., Mamajek, E.~E., \& Naylor, T. 2015, Mon. Not. R. Astron. Soc.,
  454, 593

\bibitem[{Binks \& Jeffries(2014)}]{Binks2014}
Binks, A.~S. \& Jeffries, R.~D. 2014, Mon. Not. R. Astron. Soc. Lett., 438, L11

\bibitem[{Binney \& Tremaine(1987)}]{Binney1987}
Binney, J. \& Tremaine, S. 1987, {Galactic dynamics} (Princeton, N.J. :
  Princeton University Press, c1987)

\bibitem[{Borucki {et~al.}(2007)Borucki, Koch, Basri, Batalha, Brown, Caldwell,
  Christensen-Dalsgaard, Cochran, Dunham, Gautier, Geary, Gilliland, Jenkins,
  Kondo, Latham, Lissauer, \& Monet}]{Borucki2007}
Borucki, W., Koch, D., Basri, G., {et~al.} 2007, Proc. Int. Astron. Union, 3,
  17

\bibitem[{Bressan {et~al.}(2012)Bressan, Marigo, Girardi, Salasnich, {Dal
  Cero}, Rubele, \& Nanni}]{Bressan2012}
Bressan, A., Marigo, P., Girardi, L., {et~al.} 2012, Mon. Not. R. Astron. Soc.,
  427, 127

\bibitem[{Brown {et~al.}(2020)Brown, Vallenari, Prusti, \&
  de~Bruijne}]{Brown2020}
Brown, A.~G., Vallenari, A., Prusti, T., \& de~Bruijne, J.~H. 2020, A{\&}A, 1

\bibitem[{Brown {et~al.}(2013)Brown, Baliber, Bianco, Bowman, Burleson, Conway,
  Crellin, Depagne, {De Vera}, Dilday, Dragomir, Dubberley, Eastman, Elphick,
  Falarski, Foale, Ford, Fulton, Garza, Gomez, Graham, Greene, Haldeman,
  Hawkins, Haworth, Haynes, Hidas, Hjelstrom, Howell, Hygelund, Lister,
  Lobdill, Martinez, Mullins, Norbury, Parrent, Paulson, Petry, Pickles,
  Posner, Rosing, Ross, Sand, Saunders, Shobbrook, Shporer, Street, Thomas,
  Tsapras, Tufts, Valenti, {Vander Horst}, Walker, White, \&
  Willis}]{Brown2013}
Brown, T.~M., Baliber, N., Bianco, F.~B., {et~al.} 2013, Publ. Astron. Soc.
  Pacific, 125, 1031

\bibitem[{Chen {et~al.}(2014)Chen, Girardi, Bressan, Marigo, Barbieri, \&
  Kong}]{Chen2014}
Chen, Y., Girardi, L., Bressan, A., {et~al.} 2014, Mon. Not. R. Astron. Soc.,
  444, 2525

\bibitem[{Choi {et~al.}(2016)Choi, Dotter, Conroy, Cantiello, Paxton, \&
  Johnson}]{Choi2016}
Choi, J., Dotter, A., Conroy, C., {et~al.} 2016, ApJ, 823, 102

\bibitem[{Cosentino {et~al.}(2014)Cosentino, Lovis, Pepe, Cameron, Latham,
  Molinari, Udry, Bezawada, Buchschacher, Figueira, Fleury, Ghedina, Glenday,
  Gonzalez, Guerra, Henry, Hughes, Maire, Motalebi, \&
  Phillips}]{Cosentino2014}
Cosentino, R., Lovis, C., Pepe, F., {et~al.} 2014, in Ground-based Airborne
  Instrum. Astron. V, ed. S.~K. Ramsay, I.~S. McLean, \& H.~Takami, 91478C

\bibitem[{Cosentino {et~al.}(2012)Cosentino, Lovis, Pepe, {Collier Cameron},
  Latham, Molinari, Udry, Bezawada, Black, Born, Buchschacher, Charbonneau,
  Figueira, Fleury, Galli, Gallie, Gao, Ghedina, Gonzalez, Gonzalez, Guerra,
  Henry, Horne, Hughes, Kelly, Lodi, Lunney, Maire, Mayor, Micela, Ordway,
  Peacock, Phillips, Piotto, Pollacco, Queloz, Rice, Riverol, Riverol, {San
  Juan}, Sasselov, Segransan, Sozzetti, Sosnowska, Stobie, Szentgyorgyi, Vick,
  \& Weber}]{Cosentino2012}
Cosentino, R., Lovis, C., Pepe, F., {et~al.} 2012, in Ground-based Airborne
  Instrum. Astron. IV, ed. I.~S. McLean, S.~K. Ramsay, \& H.~Takami No.
  September, 84461V

\bibitem[{Dahm(2015)}]{Dahm2015}
Dahm, S.~E. 2015, ApJ, 813, 108

\bibitem[{David {et~al.}(2016)David, Conroy, Hillenbrand, Stassun, Stauffer,
  Rebull, Cody, Isaacson, Howard, \& Aigrain}]{David2016}
David, T.~J., Conroy, K.~E., Hillenbrand, L.~A., {et~al.} 2016, Astron. J.,
  151, 112

\bibitem[{David {et~al.}(2019)David, Hillenbrand, Gillen, Cody, Howell,
  Isaacson, \& Livingston}]{David2019}
David, T.~J., Hillenbrand, L.~A., Gillen, E., {et~al.} 2019, ApJ, 872, 161

\bibitem[{Dias {et~al.}(2002)Dias, Alessi, Moitinho, \&
  L{\'{e}}pine}]{Dias2002}
Dias, W.~S., Alessi, B.~S., Moitinho, A., \& L{\'{e}}pine, J. R.~D. 2002,
  A{\&}A, 389, 871

\bibitem[{Dobbie {et~al.}(2010)Dobbie, Lodieu, \& Sharp}]{Dobbie2010}
Dobbie, P.~D., Lodieu, N., \& Sharp, R.~G. 2010, Mon. Not. R. Astron. Soc.,
  409, 1002

\bibitem[{Dotter(2016)}]{Dotter2016}
Dotter, A. 2016, Astrophys. J. Suppl. Ser., 222, 8

\bibitem[{Douglas {et~al.}(2016)Douglas, Ag{\"{u}}eros, Covey, Cargile,
  Barclay, Cody, Howell, \& Kopytova}]{Douglas2016}
Douglas, S.~T., Ag{\"{u}}eros, M.~A., Covey, K.~R., {et~al.} 2016, ApJ, 822, 47

\bibitem[{Douglas {et~al.}(2019)Douglas, Curtis, Ag{\"{u}}eros, Cargile,
  Brewer, Meibom, \& Jansen}]{Douglas2019}
Douglas, S.~T., Curtis, J.~L., Ag{\"{u}}eros, M.~A., {et~al.} 2019, ApJ, 879,
  100

\bibitem[{Dudorov \& Eretnova(2016)}]{Dudorov2016}
Dudorov, A.~E. \& Eretnova, O.~V. 2016, Astron. Astrophys. Trans., 29, 437

\bibitem[{Eggen(1975)}]{Eggen1975}
Eggen, O.~J. 1975, Publ. Astron. Soc. Pacific, 87, 37

\bibitem[{Elliott {et~al.}(2016)Elliott, Bayo, Melo, Torres, Sterzik, Quast,
  Montes, \& Brahm}]{Elliott2016}
Elliott, P., Bayo, A., Melo, C. H.~F., {et~al.} 2016, A{\&}A, 590, A13

\bibitem[{Fang {et~al.}(2018)Fang, Zhao, Zhao, \& {Bharat Kumar}}]{Fang2018}
Fang, X.-S., Zhao, G., Zhao, J.-K., \& {Bharat Kumar}, Y. 2018, Mon. Not. R.
  Astron. Soc., 476, 908

\bibitem[{Findeisen {et~al.}(2011)Findeisen, Hillenbrand, \&
  Soderblom}]{Findeisen2011}
Findeisen, K., Hillenbrand, L., \& Soderblom, D. 2011, Astron. J., 142, 23

\bibitem[{Fleming {et~al.}(2019)Fleming, Barnes, Davenport, \&
  Luger}]{Fleming2019}
Fleming, D.~P., Barnes, R., Davenport, J. R.~A., \& Luger, R. 2019, ApJ, 881,
  88

\bibitem[{Gagn{\'{e}} {et~al.}(2018)Gagn{\'{e}}, Mamajek, Malo, Riedel,
  Rodriguez, Lafreni{\`{e}}re, Faherty, Roy-Loubier, Pueyo, Robin, \&
  Doyon}]{Gagne2018}
Gagn{\'{e}}, J., Mamajek, E.~E., Malo, L., {et~al.} 2018, ApJ, 856, 23

\bibitem[{Galicher {et~al.}(2016)Galicher, Marois, Macintosh, Zuckerman,
  Barman, Konopacky, Song, Patience, Lafreni{\`{e}}re, Doyon, \&
  Nielsen}]{Galicher2016}
Galicher, R., Marois, C., Macintosh, B., {et~al.} 2016, A{\&}A, 594, A63

\bibitem[{Gillen {et~al.}(2017)Gillen, Hillenbrand, David, Aigrain, Rebull,
  Stauffer, Cody, \& Queloz}]{Gillen2017}
Gillen, E., Hillenbrand, L.~A., David, T.~J., {et~al.} 2017, ApJ, 849, 11

\bibitem[{Gizis {et~al.}(2002)Gizis, Reid, \& Hawley}]{Gizis2002}
Gizis, J.~E., Reid, I.~N., \& Hawley, S.~L. 2002, Astron. J., 123, 3356

\bibitem[{Gliese \& Jahreiss(1991)}]{Gliese1991}
Gliese, W. \& Jahreiss, H. 1991, NASA STI/Recon Tech. Rep. A, 224, 161

\bibitem[{{Guti{\'{e}}rrez Albarr{\'{a}}n} {et~al.}(2020){Guti{\'{e}}rrez
  Albarr{\'{a}}n}, Montes, {G{\'{o}}mez Garrido}, Tabernero, {Gonz{\'{a}}lez
  Hern{\'{a}}ndez}, Marfil, Frasca, Lanzafame, Klutsch, Franciosini, Randich,
  Smiljanic, Korn, Gilmore, Alfaro, Baratella, Bayo, Bensby, Bonito, Carraro,
  {Delgado Mena}, Feltzing, Gonneau, Heiter, Hourihane, {Jim{\'{e}}nez
  Esteban}, Jofre, Masseron, Monaco, Morbidelli, Prisinzano, Roccatagliata,
  Sousa, {Van der Swaelmen}, Worley, \& Zaggia}]{GutierrezAlbarran2020}
{Guti{\'{e}}rrez Albarr{\'{a}}n}, M.~L., Montes, D., {G{\'{o}}mez Garrido}, M.,
  {et~al.} 2020, A{\&}A, 643, A71

\bibitem[{Hawley {et~al.}(1996)Hawley, Gizis, \& Reid}]{Hawley1996}
Hawley, S.~L., Gizis, J.~E., \& Reid, I.~N. 1996, Astron. J., 112, 2799

\bibitem[{Hebb {et~al.}(2006)Hebb, Wyse, Gilmore, \& Holtzman}]{Hebb2006}
Hebb, L., Wyse, R. F.~G., Gilmore, G., \& Holtzman, J. 2006, Astron. J., 131,
  555

\bibitem[{Heintz(1986{\natexlab{a}})}]{Heintz1986b}
Heintz, W.~D. 1986{\natexlab{a}}, Astron. Astrophys. Suppl. Ser., 65, 411

\bibitem[{Heintz(1986{\natexlab{b}})}]{Heintz1986a}
Heintz, W.~D. 1986{\natexlab{b}}, Astron. J., 92, 446

\bibitem[{Henden {et~al.}(2016)Henden, Templeton, Terrell, Smith, Levine, \&
  Welch}]{Henden2016}
Henden, A.~A., Templeton, M., Terrell, D., {et~al.} 2016, VizieR Online Data
  Cat., II/336

\bibitem[{Henry {et~al.}(2018)Henry, Jao, Winters, Dieterich, Finch, Ianna,
  Riedel, Silverstein, Subasavage, \& Vrijmoet}]{Henry2018}
Henry, T.~J., Jao, W.-C., Winters, J.~G., {et~al.} 2018, Astron. J., 155, 265

\bibitem[{Herczeg \& Hillenbrand(2015)}]{Herczeg2015}
Herczeg, G.~J. \& Hillenbrand, L.~A. 2015, ApJ, 808, 23

\bibitem[{Husser {et~al.}(2013)Husser, {Wende-von Berg}, Dreizler, Homeier,
  Reiners, Barman, \& Hauschildt}]{Husser2013}
Husser, T.-O., {Wende-von Berg}, S., Dreizler, S., {et~al.} 2013, A{\&}A, 553,
  A6

\bibitem[{Janson {et~al.}(2017)Janson, Durkan, Hippler, Dai, Brandner,
  Schlieder, Bonnefoy, \& Henning}]{Janson2017}
Janson, M., Durkan, S., Hippler, S., {et~al.} 2017, A{\&}A, 599, A70

\bibitem[{Jeffers {et~al.}(2018)Jeffers, Sch{\"{o}}fer, Lamert, Reiners,
  Montes, Caballero, Cort{\'{e}}s-Contreras, Marvin, Passegger, Zechmeister,
  Quirrenbach, Alonso-Floriano, Amado, Bauer, Casal, {Diez Alonso}, Herrero,
  Morales, Mundt, Ribas, \& Sarmiento}]{Jeffers2018}
Jeffers, S.~V., Sch{\"{o}}fer, P., Lamert, A., {et~al.} 2018, A{\&}A, 614, A76

\bibitem[{Johnson \& Soderblom(1987)}]{Johnson1987}
Johnson, D. R.~H. \& Soderblom, D.~R. 1987, Astron. J., 93, 864

\bibitem[{Jones \& West(2016)}]{Jones2016}
Jones, D.~O. \& West, A.~A. 2016, ApJ, 817, 1

\bibitem[{Koch {et~al.}(2004)Koch, Borucki, Dunham, Geary, Gilliland, Jenkins,
  Latham, Bachtell, Berry, Deininger, Duren, Gautier, Gillis, Mayer, Miller,
  Shafer, Sobeck, Stewart, \& Weiss}]{Koch2004}
Koch, D.~G., Borucki, W., Dunham, E., {et~al.} 2004, 1491

\bibitem[{Kraus {et~al.}(2017)Kraus, Douglas, Mann, Ag{\"{u}}eros, Law, Covey,
  Feiden, Rizzuto, Howard, Isaacson, Gaidos, Torres, \& Bakos}]{Kraus2017}
Kraus, A.~L., Douglas, S.~T., Mann, A.~W., {et~al.} 2017, ApJ, 845, 72

\bibitem[{Kuiper(1943)}]{Kuiper1943}
Kuiper, V. G.~P. 1943, ApJ, 97, 275

\bibitem[{Lodieu(2020)}]{Lodieu2020b}
Lodieu, N. 2020, Mem. della Soc. Astron. Ital., 91, 84

\bibitem[{Lodieu {et~al.}(2020)Lodieu, Paunzen, \& Zejda}]{Lodieu2020a}
Lodieu, N., Paunzen, E., \& Zejda, M. 2020, in Rev. Front. Mod. Astrophys.
  (Cham: Springer International Publishing), 213--243

\bibitem[{L{\'{o}}pez-Santiago {et~al.}(2009)L{\'{o}}pez-Santiago, Micela, \&
  Montes}]{Lopez-Santiago2009}
L{\'{o}}pez-Santiago, J., Micela, G., \& Montes, D. 2009, A{\&}A, 499, 129

\bibitem[{L{\'{o}}pez-Santiago {et~al.}(2006)L{\'{o}}pez-Santiago, Montes,
  Crespo‐Chacon, \& Fernandez‐Figueroa}]{LopezSantiago2006}
L{\'{o}}pez-Santiago, J., Montes, D., Crespo‐Chacon, I., \&
  Fernandez‐Figueroa, M.~J. 2006, ApJ, 643, 1160

\bibitem[{L{\'{o}}pez-Santiago {et~al.}(2010)L{\'{o}}pez-Santiago, Montes,
  G{\'{a}}lvez-Ortiz, Crespo-Chac{\'{o}}n, Mart{\'{i}}nez-Arn{\'{a}}iz,
  Fern{\'{a}}ndez-Figueroa, de~Castro, \& Cornide}]{Lopez-Santiago2010}
L{\'{o}}pez-Santiago, J., Montes, D., G{\'{a}}lvez-Ortiz, M.~C., {et~al.} 2010,
  A{\&}A, 514, A97

\bibitem[{Makarov {et~al.}(2008)Makarov, Zacharias, \& Hennessy}]{Makarov2008}
Makarov, V.~V., Zacharias, N., \& Hennessy, G.~S. 2008, ApJ, 687, 566

\bibitem[{Malo {et~al.}(2014{\natexlab{a}})Malo, Artigau, Doyon,
  Lafreni{\`{e}}re, Albert, \& Gagn{\'{e}}}]{Malo2014a}
Malo, L., Artigau, {\'{E}}., Doyon, R., {et~al.} 2014{\natexlab{a}}, ApJ, 788,
  81

\bibitem[{Malo {et~al.}(2014{\natexlab{b}})Malo, Doyon, Feiden, Albert,
  Lafreni{\`{e}}re, Artigau, Gagn{\'{e}}, \& Riedel}]{Malo2014b}
Malo, L., Doyon, R., Feiden, G.~A., {et~al.} 2014{\natexlab{b}}, ApJ, 792, 37

\bibitem[{Malo {et~al.}(2013)Malo, Doyon, Lafreni{\`{e}}re, Artigau,
  Gagn{\'{e}}, Baron, \& Riedel}]{Malo2013}
Malo, L., Doyon, R., Lafreni{\`{e}}re, D., {et~al.} 2013, ApJ, 762, 88

\bibitem[{Martin {et~al.}(2005)Martin, Fanson, Schiminovich, Morrissey,
  Friedman, Barlow, Conrow, Grange, Jelinsky, Milliard, Siegmund, Bianchi,
  Byun, Donas, Forster, Heckman, Lee, Madore, Malina, Neff, Rich, Small,
  Surber, Szalay, Welsh, \& Wyder}]{Martin2005}
Martin, D.~C., Fanson, J., Schiminovich, D., {et~al.} 2005, ApJ, 619, L1

\bibitem[{Mart{\'{i}}n {et~al.}(2018)Mart{\'{i}}n, Lodieu, Pavlenko, \&
  B{\'{e}}jar}]{Martin2018}
Mart{\'{i}}n, E.~L., Lodieu, N., Pavlenko, Y., \& B{\'{e}}jar, V. J.~S. 2018,
  ApJ, 856, 40

\bibitem[{Mason {et~al.}(2002)Mason, Hartkopf, Urban, Wycoff, Pascu, Hall,
  Hennessy, Platais, Rafferty, \& Holdenried}]{Mason2002}
Mason, B.~D., Hartkopf, W.~I., Urban, S.~E., {et~al.} 2002, Astron. J., 124,
  2254

\bibitem[{Mayor {et~al.}(2003)Mayor, Pepe, Queloz, Bouchy, Rupprecht, {Lo
  Curto}, Avila, Benz, Bertaux, Bonfils, Dall, Dekker, Delabre, Eckert, Fleury,
  Gilliotte, Gojak, Guzman, Kohler, Lizon, Longinotti, Lovis, Megevand,
  Pasquini, Reyes, Sivan, Sosnowska, Soto, Udry, van Kesteren, Weber, \&
  Weilenmann}]{Mayor2003}
Mayor, M., Pepe, F., Queloz, D., {et~al.} 2003, The Messenger, 114, 20

\bibitem[{Meibom \& Mathieu(2005)}]{Meibom2005}
Meibom, S. \& Mathieu, R.~D. 2005, ApJ, 620, 970

\bibitem[{Meshkat {et~al.}(2017)Meshkat, Mawet, Bryan, Hinkley, Bowler,
  Stapelfeldt, Batygin, Padgett, Morales, Serabyn, Christiaens, Brandt, \&
  Wahhaj}]{Meshkat2017}
Meshkat, T., Mawet, D., Bryan, M.~L., {et~al.} 2017, Astron. J., 154, 245

\bibitem[{Montes {et~al.}(2001)Montes, L{\'{o}}pez-Santiago, G{\'{a}}lvez,
  Fern{\'{a}}ndez-Figueroa, {De Castro}, \& Cornide}]{Montes2001}
Montes, D., L{\'{o}}pez-Santiago, J., G{\'{a}}lvez, M., {et~al.} 2001, Mon.
  Not. R. Astron. Soc., 328, 45

\bibitem[{Myers {et~al.}(2015)Myers, Sande, Miller, {Warren, W. H.}, \&
  Tracewell}]{Myers2015}
Myers, J.~R., Sande, C.~B., Miller, A.~C., {Warren, W. H.}, J., \& Tracewell,
  D.~A. 2015, VizieR Online Data Cat., V/145

\bibitem[{Naud {et~al.}(2017)Naud, Artigau, Doyon, Malo, Gagn{\'{e}},
  Lafreni{\`{e}}re, Wolf, \& Magnier}]{Naud2017}
Naud, M.-E., Artigau, {\'{E}}., Doyon, R., {et~al.} 2017, Astron. J., 154, 129

\bibitem[{Nielsen {et~al.}(2019)Nielsen, {De Rosa}, Macintosh, Wang, Ruffio,
  Chiang, Marley, Saumon, Savransky, {Mark Ammons}, Bailey, Barman, Blain,
  Bulger, Burrows, Chilcote, Cotten, Czekala, Doyon, Duch{\^{e}}ne, Esposito,
  Fabrycky, Fitzgerald, Follette, Fortney, Gerard, Goodsell, Graham, Greenbaum,
  Hibon, Hinkley, Hirsch, Hom, Hung, {Ilene Dawson}, Ingraham, Kalas,
  Konopacky, Larkin, Lee, Lin, Maire, Marchis, Marois, Metchev,
  Millar-Blanchaer, Morzinski, Oppenheimer, Palmer, Patience, Perrin, Poyneer,
  Pueyo, Rafikov, Rajan, Rameau, Rantakyr{\"{o}}, Ren, Schneider,
  Sivaramakrishnan, Song, Soummer, Tallis, Thomas, Ward-Duong, \&
  Wolff}]{Nielsen2019}
Nielsen, E.~L., {De Rosa}, R.~J., Macintosh, B., {et~al.} 2019, Astron. J.,
  158, 13

\bibitem[{N{\'{u}}{\~{n}}ez \& Ag{\"{u}}eros(2016)}]{Nunez2016}
N{\'{u}}{\~{n}}ez, A. \& Ag{\"{u}}eros, M.~A. 2016, ApJ, 830, 44

\bibitem[{Pecaut \& Mamajek(2013)}]{Pecaut2013}
Pecaut, M.~J. \& Mamajek, E.~E. 2013, Astrophys. J. Suppl. Ser., 208, 9

\bibitem[{Piskunov \& Valenti(2002)}]{Piskunov2002}
Piskunov, N.~E. \& Valenti, J.~A. 2002, A{\&}A, 385, 1095

\bibitem[{Pollacco {et~al.}(2006)Pollacco, Skillen, Cameron, Christian,
  Hellier, Irwin, Lister, Street, West, Anderson, Clarkson, Deeg, Enoch, Evans,
  Fitzsimmons, Haswell, Hodgkin, Horne, Kane, Keenan, Maxted, Norton, Osborne,
  Parley, Ryans, Smalley, Wheatley, \& Wilson}]{Pollacco2006}
Pollacco, D.~L., Skillen, I., Cameron, A.~C., {et~al.} 2006, Publ. Astron. Soc.
  Pacific, 118, 1407

\bibitem[{Pourbaix {et~al.}(2004)Pourbaix, Tokovinin, Batten, Fekel, Hartkopf,
  Levato, Morrell, Torres, \& Udry}]{Pourbaix2004}
Pourbaix, D., Tokovinin, A.~A., Batten, A.~H., {et~al.} 2004, A{\&}A, 424, 727

\bibitem[{Prusti {et~al.}(2016)Prusti, de~Bruijne, Brown, Vallenari, Babusiaux,
  Bailer-Jones, Bastian, Biermann, Evans, Eyer, Jansen, Jordi, Klioner,
  Lammers, Lindegren, Luri, Mignard, Milligan, Panem, Poinsignon, Pourbaix,
  Randich, Sarri, Sartoretti, Siddiqui, Soubiran, Valette, van Leeuwen, Walton,
  Aerts, Arenou, Cropper, Drimmel, H{\o}g, Katz, Lattanzi, O'Mullane, Grebel,
  Holland, Huc, Passot, Bramante, Cacciari, Casta{\~{n}}eda, Chaoul, Cheek, {De
  Angeli}, Fabricius, Guerra, Hern{\'{a}}ndez, Jean-Antoine-Piccolo, Masana,
  Messineo, Mowlavi, Nienartowicz, Ord{\'{o}}{\~{n}}ez-Blanco, Panuzzo,
  Portell, Richards, Riello, Seabroke, Tanga, Th{\'{e}}venin, Torra, Els,
  Gracia-Abril, Comoretto, Garcia-Reinaldos, Lock, Mercier, Altmann, Andrae,
  Astraatmadja, Bellas-Velidis, Benson, Berthier, Blomme, Busso, Carry,
  Cellino, Clementini, Cowell, Creevey, Cuypers, Davidson, {De Ridder},
  de~Torres, Delchambre, Dell'Oro, Ducourant, Fr{\'{e}}mat,
  Garc{\'{i}}a-Torres, Gosset, Halbwachs, Hambly, Harrison, Hauser, Hestroffer,
  Hodgkin, Huckle, Hutton, Jasniewicz, Jordan, Kontizas, Korn, Lanzafame,
  Manteiga, Moitinho, Muinonen, Osinde, Pancino, Pauwels, Petit, Recio-Blanco,
  Robin, Sarro, Siopis, Smith, Smith, Sozzetti, Thuillot, van Reeven, Viala,
  Abbas, {Abreu Aramburu}, Accart, Aguado, Allan, Allasia, Altavilla,
  {\'{A}}lvarez, Alves, Anderson, Andrei, {Anglada Varela}, Antiche, Antoja,
  Ant{\'{o}}n, Arcay, Atzei, Ayache, Bach, Baker, Balaguer-N{\'{u}}{\~{n}}ez,
  Barache, Barata, Barbier, Barblan, Baroni, {Barrado y Navascu{\'{e}}s},
  Barros, Barstow, Becciani, Bellazzini, Bellei, {Bello Garc{\'{i}}a},
  Belokurov, Bendjoya, Berihuete, Bianchi, Bienaym{\'{e}}, Billebaud,
  Blagorodnova, Blanco-Cuaresma, Boch, Bombrun, Borrachero, Bouquillon, Bourda,
  Bouy, Bragaglia, Breddels, Brouillet, Br{\"{u}}semeister, Bucciarelli,
  Budnik, Burgess, Burgon, Burlacu, Busonero, Buzzi, Caffau, Cambras, Campbell,
  Cancelliere, Cantat-Gaudin, Carlucci, Carrasco, Castellani, Charlot, Charnas,
  Charvet, Chassat, Chiavassa, Clotet, Cocozza, Collins, Collins, Costigan,
  Crifo, Cross, Crosta, Crowley, Dafonte, Damerdji, Dapergolas, David, David,
  {De Cat}, de~Felice, de~Laverny, {De Luise}, {De March}, de~Martino,
  de~Souza, Debosscher, del Pozo, Delbo, Delgado, Delgado, di~Marco, {Di
  Matteo}, Diakite, Distefano, Dolding, {Dos Anjos}, Drazinos, Dur{\'{a}}n,
  Dzigan, Ecale, Edvardsson, Enke, Erdmann, Escolar, Espina, Evans, {Eynard
  Bontemps}, Fabre, Fabrizio, Faigler, Falc{\~{a}}o, {Farr{\`{a}}s Casas},
  Faye, Federici, Fedorets, Fern{\'{a}}ndez-Hern{\'{a}}ndez, Fernique, Fienga,
  Figueras, Filippi, Findeisen, Fonti, Fouesneau, Fraile, Fraser, Fuchs,
  Furnell, Gai, Galleti, Galluccio, Garabato, Garc{\'{i}}a-Sedano, Gar{\'{e}},
  Garofalo, Garralda, Gavras, Gerssen, Geyer, Gilmore, Girona, Giuffrida,
  Gomes, Gonz{\'{a}}lez-Marcos, Gonz{\'{a}}lez-N{\'{u}}{\~{n}}ez,
  Gonz{\'{a}}lez-Vidal, Granvik, Guerrier, Guillout, Guiraud, G{\'{u}}rpide,
  Guti{\'{e}}rrez-S{\'{a}}nchez, Guy, Haigron, Hatzidimitriou, Haywood, Heiter,
  Helmi, Hobbs, Hofmann, Holl, Holland, Hunt, Hypki, Icardi, Irwin, {Jevardat
  de Fombelle}, Jofr{\'{e}}, Jonker, Jorissen, Julbe, Karampelas, Kochoska,
  Kohley, Kolenberg, Kontizas, Koposov, Kordopatis, Koubsky, Kowalczyk,
  Krone-Martins, Kudryashova, Kull, Bachchan, Lacoste-Seris, Lanza, Lavigne,
  {Le Poncin-Lafitte}, Lebreton, Lebzelter, Leccia, Leclerc, Lecoeur-Taibi,
  Lemaitre, Lenhardt, Leroux, Liao, Licata, Lindstr{\o}m, Lister, Livanou,
  Lobel, L{\"{o}}ffler, L{\'{o}}pez, Lopez-Lozano, Lorenz, Loureiro, MacDonald,
  {Magalh{\~{a}}es Fernandes}, Managau, Mann, Mantelet, Marchal, Marchant,
  Marconi, Marie, Marinoni, Marrese, Marschalk{\'{o}}, Marshall,
  Mart{\'{i}}n-Fleitas, Martino, Mary, Matijevi{\v{c}}, Mazeh, McMillan,
  Messina, Mestre, Michalik, Millar, Miranda, Molina, Molinaro, Molinaro,
  Moln{\'{a}}r, Moniez, Montegriffo, Monteiro, Mor, Mora, Morbidelli, Morel,
  Morgenthaler, Morley, Morris, Mulone, Muraveva, Musella, Narbonne, Nelemans,
  Nicastro, Noval, Ord{\'{e}}novic, Ordieres-Mer{\'{e}}, Osborne, Pagani,
  Pagano, Pailler, Palacin, Palaversa, Parsons, Paulsen, Pecoraro, Pedrosa,
  Pentik{\"{a}}inen, Pereira, Pichon, Piersimoni, Pineau, Plachy, Plum,
  Poujoulet, Pr{\v{s}}a, Pulone, Ragaini, Rago, Rambaux, Ramos-Lerate, Ranalli,
  Rauw, Read, Regibo, Renk, Reyl{\'{e}}, Ribeiro, Rimoldini, Ripepi, Riva,
  Rixon, Roelens, Romero-G{\'{o}}mez, Rowell, Royer, Rudolph, Ruiz-Dern,
  Sadowski, {Sagrist{\`{a}} Sell{\'{e}}s}, Sahlmann, Salgado, Salguero,
  Sarasso, Savietto, Schnorhk, Schultheis, Sciacca, Segol, Segovia, Segransan,
  Serpell, Shih, Smareglia, Smart, Smith, Solano, Solitro, Sordo, {Soria
  Nieto}, Souchay, Spagna, Spoto, Stampa, Steele, Steidelm{\"{u}}ller,
  Stephenson, Stoev, Suess, S{\"{u}}veges, Surdej, Szabados, Szegedi-Elek,
  Tapiador, Taris, Tauran, Taylor, Teixeira, Terrett, Tingley, Trager, Turon,
  Ulla, Utrilla, Valentini, van Elteren, {Van Hemelryck}, van Leeuwen, Varadi,
  Vecchiato, Veljanoski, Via, Vicente, Vogt, Voss, Votruba, Voutsinas,
  Walmsley, Weiler, Weingrill, Werner, Wevers, Whitehead, Wyrzykowski, Yoldas,
  {\v{Z}}erjal, Zucker, Zurbach, Zwitter, Alecu, Allen, {Allende Prieto},
  Amorim, Anglada-Escud{\'{e}}, Arsenijevic, Azaz, Balm, Beck, Bernstein,
  Bigot, Bijaoui, Blasco, Bonfigli, Bono, Boudreault, Bressan, Brown, Brunet,
  Bunclark, Buonanno, Butkevich, Carret, Carrion, Chemin, Ch{\'{e}}reau,
  Corcione, Darmigny, de~Boer, de~Teodoro, de~Zeeuw, {Delle Luche}, Domingues,
  Dubath, Fodor, Fr{\'{e}}zouls, Fries, Fustes, Fyfe, Gallardo, Gallegos,
  Gardiol, Gebran, Gomboc, G{\'{o}}mez, Grux, Gueguen, Heyrovsky, Hoar,
  Iannicola, {Isasi Parache}, Janotto, Joliet, Jonckheere, Keil, Kim,
  Klagyivik, Klar, Knude, Kochukhov, Kolka, Kos, Kutka, Lainey, LeBouquin, Liu,
  Loreggia, Makarov, Marseille, Martayan, Martinez-Rubi, Massart, Meynadier,
  Mignot, Munari, Nguyen, Nordlander, Ocvirk, O'Flaherty, {Olias Sanz}, Ortiz,
  Osorio, Oszkiewicz, Ouzounis, Palmer, Park, Pasquato, Peltzer, Peralta,
  P{\'{e}}turaud, Pieniluoma, Pigozzi, Poels, Prat, Prod'homme, Raison,
  Rebordao, Risquez, Rocca-Volmerange, Rosen, Ruiz-Fuertes, Russo, Sembay,
  {Serraller Vizcaino}, Short, Siebert, Silva, Sinachopoulos, Slezak, Soffel,
  Sosnowska, Strai{\v{z}}ys, ter Linden, Terrell, Theil, Tiede, Troisi,
  Tsalmantza, Tur, Vaccari, Vachier, Valles, {Van Hamme}, Veltz, Virtanen,
  Wallut, Wichmann, Wilkinson, Ziaeepour, \& Zschocke}]{Prusti2016}
Prusti, T., de~Bruijne, J. H.~J., Brown, A. G.~A., {et~al.} 2016, A{\&}A, 595,
  A1

\bibitem[{Raskin {et~al.}(2011)Raskin, {Van Winckel}, Hensberge, Jorissen,
  Lehmann, Waelkens, Avila, \& {De Cuyper}}]{Raskin2011}
Raskin, G., {Van Winckel}, H., Hensberge, H., {et~al.} 2011, A{\&}A, 69, 1

\bibitem[{Rebull {et~al.}(2016)Rebull, Stauffer, Bouvier, Cody, Hillenbrand,
  Soderblom, Valenti, Barrado, Bouy, Ciardi, Pinsonneault, Stassun, Micela,
  Aigrain, Vrba, Somers, Christiansen, Gillen, \& Cameron}]{Rebull2016}
Rebull, L.~M., Stauffer, J.~R., Bouvier, J., {et~al.} 2016, Astron. J., 152,
  113

\bibitem[{Riaz {et~al.}(2006)Riaz, Gizis, \& Harvin}]{Riaz2006}
Riaz, B., Gizis, J.~E., \& Harvin, J. 2006, Astron. J., 132, 866

\bibitem[{Ricker {et~al.}(2014)Ricker, Winn, Vanderspek, Latham, Bakos, Bean,
  Berta-Thompson, Brown, Buchhave, Butler, Butler, Chaplin, Charbonneau,
  Christensen-Dalsgaard, Clampin, Deming, Doty, {De Lee}, Dressing, Dunham,
  Endl, Fressin, Ge, Henning, Holman, Howard, Ida, Jenkins, Jernigan, Johnson,
  Kaltenegger, Kawai, Kjeldsen, Laughlin, Levine, Lin, Lissauer, MacQueen,
  Marcy, McCullough, Morton, Narita, Paegert, Palle, Pepe, Pepper, Quirrenbach,
  Rinehart, Sasselov, Sato, Seager, Sozzetti, Stassun, Sullivan, Szentgyorgyi,
  Torres, Udry, \& Villase{\~{n}}or}]{Ricker2014}
Ricker, G.~R., Winn, J.~N., Vanderspek, R., {et~al.} 2014, J. Astron. Telesc.
  Instruments, Syst., 1, 014003

\bibitem[{Riedel {et~al.}(2017)Riedel, Blunt, Lambrides, Rice, Cruz, \&
  Faherty}]{Riedel2017}
Riedel, A.~R., Blunt, S.~C., Lambrides, E.~L., {et~al.} 2017, Astron. J., 153,
  95

\bibitem[{Schmitt {et~al.}(1995)Schmitt, Fleming, \& Giampapa}]{Schmitt1995}
Schmitt, J. H. M.~M., Fleming, T.~A., \& Giampapa, M.~S. 1995, ApJ, 450, 392

\bibitem[{Schneider {et~al.}(2019)Schneider, Shkolnik, Allers, Kraus, Liu,
  Weinberger, \& Flagg}]{Schneider2019}
Schneider, A.~C., Shkolnik, E.~L., Allers, K.~N., {et~al.} 2019, Astron. J.,
  157, 234

\bibitem[{Shaya \& Olling(2011)}]{Shaya2011}
Shaya, E.~J. \& Olling, R.~P. 2011, Astrophys. J. Suppl. Ser., 192, 2

\bibitem[{Shkolnik {et~al.}(2017)Shkolnik, Allers, Kraus, Liu, \&
  Flagg}]{Shkolnik2017}
Shkolnik, E.~L., Allers, K.~N., Kraus, A.~L., Liu, M.~C., \& Flagg, L. 2017,
  Astron. J., 154, 69

\bibitem[{Siverd {et~al.}(2018)Siverd, Brown, Henderson, Hygelund, Barnes, {De
  Vera}, Kirby, Smith, Taylor, Tufts, Bowman, Foale, McCully, Nation, \&
  Harbeck}]{Siverd2018}
Siverd, R.~J., Brown, T., Henderson, T., {et~al.} 2018, in Ground-based
  Airborne Instrum. Astron. VII, ed. H.~Takami, C.~J. Evans, \& L.~Simard No.
  July 2018 (SPIE), 231

\bibitem[{Skrutskie {et~al.}(2006)Skrutskie, Cutri, Stiening, Weinberg,
  Schneider, Carpenter, Beichman, Capps, Chester, Elias, Huchra, Liebert,
  Lonsdale, Monet, Price, Seitzer, Jarrett, Kirkpatrick, Gizis, Howard, Evans,
  Fowler, Fullmer, Hurt, Light, Kopan, Marsh, McCallon, Tam, {Van Dyk}, \&
  Wheelock}]{Skrutskie2006}
Skrutskie, M.~F., Cutri, R.~M., Stiening, R., {et~al.} 2006, Astron. J., 131,
  1163

\bibitem[{Smareglia {et~al.}(2014)Smareglia, Bignamini, Knapic, Molinaro, \&
  Collaboration}]{Smareglia2014}
Smareglia, R., Bignamini, A., Knapic, C., Molinaro, M., \& Collaboration, G.
  2014, in Astron. Data Anal. Softw. Syst. XXIII, 435

\bibitem[{Soderblom(2010)}]{Soderblom2010}
Soderblom, D.~R. 2010, Annu. Rev. Astron. Astrophys., 48, 581

\bibitem[{Somers {et~al.}(2020)Somers, Cao, \& Pinsonneault}]{Somers2020}
Somers, G., Cao, L., \& Pinsonneault, M.~H. 2020, ApJ, 891, 29

\bibitem[{Somers \& Pinsonneault(2015)}]{Somers2015}
Somers, G. \& Pinsonneault, M.~H. 2015, ApJ, 807, 174

\bibitem[{Southworth(2015)}]{Southworth2015}
Southworth, J. 2015in , 164--165

\bibitem[{Sperauskas {et~al.}(2019)Sperauskas, Deveikis, \&
  Tokovinin}]{Sperauskas2019}
Sperauskas, J., Deveikis, V., \& Tokovinin, A. 2019, A{\&}A, 31, 1

\bibitem[{Stauffer {et~al.}(1998)Stauffer, Schultz, \&
  Kirkpatrick}]{Stauffer1998}
Stauffer, J.~R., Schultz, G., \& Kirkpatrick, J.~D. 1998, ApJ, 499, L199

\bibitem[{Tody(1986)}]{Tody1986}
Tody, D. 1986, 733

\bibitem[{Tokovinin(2018)}]{Tokovinin2018}
Tokovinin, A. 2018, Astrophys. J. Suppl. Ser., 235, 6

\bibitem[{Torres {et~al.}(2006)Torres, Quast, da~Silva, de~la Reza, Melo, \&
  Sterzik}]{Torres2006}
Torres, C. A.~O., Quast, G.~R., da~Silva, L., {et~al.} 2006, A{\&}A, 460, 695

\bibitem[{Torres {et~al.}(2008)Torres, Quast, Melo, \& Sterzik}]{Torres2008}
Torres, C. A.~O., Quast, G.~R., Melo, C. H.~F., \& Sterzik, M.~F. 2008, in
  Handb. Star Form. Reg. Vol. II, ed. B.~Reipurth, Vol.~5, 757

\bibitem[{Torres {et~al.}(2010)Torres, Andersen, \& Gim{\'{e}}nez}]{Torres2010}
Torres, G., Andersen, J., \& Gim{\'{e}}nez, A. 2010, Astron. Astrophys. Rev.,
  18, 67

\bibitem[{van Leeuwen(2007)}]{VanLeeuwen2007}
van Leeuwen, F. 2007, A{\&}A, 474, 653

\bibitem[{Voges {et~al.}(1999)Voges, Aschenbach, Boller, Br{\"{a}}uninger,
  Briel, Burkert, Dennerl, Englhauser, Gruber, Haberl, Hartner, Hasinger,
  K{\"{u}}rster, Pfeffermann, Pietsch, Predehl, Rosso, Schmitt, Tr{\"{u}}mper,
  \& Zimmermann}]{Voges1999}
Voges, W., Aschenbach, B., Boller, T., {et~al.} 1999, ApJ, 349, 389

\bibitem[{Weinberg {et~al.}(1987)Weinberg, Shapiro, \&
  Wasserman}]{Weinberg1987}
Weinberg, M.~D., Shapiro, S.~L., \& Wasserman, I. 1987, ApJ, 312, 367

\bibitem[{Woolley {et~al.}(1970)Woolley, Epps, Penston, \&
  Pocock}]{Woolley1970}
Woolley, R., Epps, E.~A., Penston, M.~J., \& Pocock, S.~B. 1970, R. Obs. Ann.,
  5, 227

\bibitem[{Wright \& Eastman(2014)}]{Wright2014}
Wright, J.~T. \& Eastman, J.~D. 2014, Publ. Astron. Soc. Pacific, 126, 838

\bibitem[{Zacharias {et~al.}(2013)Zacharias, Finch, Girard, Henden, Bartlett,
  Monet, \& Zacharias}]{Zacharias2013}
Zacharias, N., Finch, C.~T., Girard, T.~M., {et~al.} 2013, Astron. J., 145
  [\eprint[arXiv]{1212.6182}]

\bibitem[{{Zapatero Osorio} {et~al.}(2007){Zapatero Osorio}, Martin, Bejar,
  Bouy, Deshpande, \& Wainscoat}]{ZapateroOsorio2007}
{Zapatero Osorio}, M.~R., Martin, E.~L., Bejar, V. J.~S., {et~al.} 2007, ApJ,
  666, 1205

\bibitem[{Zechmeister \& K{\"{u}}rster(2009)}]{Zechmeister2009}
Zechmeister, M. \& K{\"{u}}rster, M. 2009, A{\&}A, 496, 577

\bibitem[{Zucker \& Mazeh(1994)}]{Zucker1994}
Zucker, S. \& Mazeh, T. 1994, ApJ, 420, 806

\bibitem[{Zucker {et~al.}(2004)Zucker, Mazeh, Santos, Udry, \&
  Mayor}]{Zucker2004}
Zucker, S., Mazeh, T., Santos, N.~C., Udry, S., \& Mayor, M. 2004, A{\&}A, 426,
  695

\bibitem[{Zuckerman \& Song(2004)}]{Zuckerman2004}
Zuckerman, B. \& Song, I. 2004, Annu. Rev. Astron. Astrophys., 42, 685

\bibitem[{Zuckerman {et~al.}(2001)Zuckerman, Song, Bessell, \&
  Webb}]{Zuckerman2001}
Zuckerman, B., Song, I., Bessell, M.~S., \& Webb, R.~A. 2001, ApJ, 562, L87

\bibitem[{Z{\'{u}}{\~{n}}iga-Fern{\'{a}}ndez
  {et~al.}(2021)Z{\'{u}}{\~{n}}iga-Fern{\'{a}}ndez, Bayo, Elliott, Zamora,
  Corval{\'{a}}n, Haubois, Corral-Santana, Olofsson, Hu{\'{e}}lamo, Sterzik,
  Torres, Quast, \& Melo}]{Zuniga-Fernandez2021}
Z{\'{u}}{\~{n}}iga-Fern{\'{a}}ndez, S., Bayo, A., Elliott, P., {et~al.} 2021,
  A{\&}A, 645, A30

\end{thebibliography}

\clearpage
\onecolumn

        \begin{longtable}{lccr}
                \caption{\label{LAProtTable}Rotation period derived from the TESS light curves of the LA members compiled from \citet{Montes2001}.}\\
                \hline\hline
                \noalign{\smallskip}
                \text{\it Gaia} eDR3 ID &      $  \alpha $ (J2000) &         $  \delta $ (J2000)&    P$ _\mathrm{rot} $ [d]  \\
                \noalign{\smallskip}
                \hline
                \endfirsthead
                \caption{continued.}\\
                \hline\hline
                \noalign{\smallskip}
                \text{\it Gaia} eDR3 ID &      $  \alpha $ (J2000) &         $  \delta $ (J2000)&    P$ _\mathrm{rot} $ [d]  \\
                \noalign{\smallskip}
                \hline
                \noalign{\smallskip}
                \endhead
                \hline
                \endfoot
                \noalign{\smallskip}
                4905600633472735232 &  00:05:28.45 &  $ - $61:13:33.06 &  13.50 \\
                4995853014646190080 &  00:05:52.55 &  $ - $41:45:11.04 &   2.95 \\
                2860924621205256704 &  00:06:36.78 &  $ + $29:01:17.41 &   6.30 \\
                2860839993168688128 &  00:18:20.90 &  $ + $30:57:22.03 &   1.77 \\
                4901229043960054784 &  00:18:26.12 &  $ - $63:28:38.97 &   2.30 \\
                4901926409210454016 &  00:24:08.98 &  $ - $62:11:04.41 &   1.75 \\
                4902014095262270464 &  00:25:14.66 &  $ - $61:30:48.33 &   1.75 \\
                532870034006715264 &  01:02:57.22 &  $ + $69:13:37.41 &  12.00 \\
                5038817840251308032 &  01:33:15.81 &  $ - $24:10:40.67 &  17.50 \\
                2477815222028038144 &  01:37:35.47 &  $ - $06:45:37.52 &   7.50 \\
                2498460442625153024 &  02:41:14.00 &  $ - $00:41:44.36 &   1.77 \\
                5160075762132997120 &  02:52:32.13 &  $ - $12:46:10.97 &   6.95 \\
                5166951386298775552 &  03:09:42.29 &  $ - $09:34:46.59 &   5.52 \\
                248004472671281664 &  03:33:13.49 &  $ + $46:15:26.54 &   3.17 \\
                3256786534197166080 &  04:02:36.74 &  $ - $00:16:08.13 &   1.50 \\
                493935296471302784 &  04:09:35.04 &  $ + $69:32:29.01 &    ... \\
                2972231722338351104 &  05:34:09.16 &  $ - $15:17:03.18 &   2.02 \\
                4795598309045006336 &  05:36:56.85 &  $ - $47:57:52.87 &   4.60 \\
                263916708025623680 &  05:41:20.34 &  $ + $53:28:51.81 &   5.53 \\
                3116883781327752704 &  06:19:08.06 &  $ - $03:26:20.38 &   1.36 \\
                5266270443442455552 &  06:18:28.21 &  $ - $72:02:41.43 &   2.67 \\
                3135225421986580992 &  07:39:23.04 &  $ + $02:11:01.18 &    ... \\
                902175202130168064 &  08:08:56.39 &  $ + $32:49:11.41 &   3.38 \\
                3097747606079523840 &  08:07:09.09 &  $ + $07:23:00.13 &  22.50 \\
                3090083696499828736 &  08:22:49.95 &  $ + $01:51:33.55 &   3.35 \\
                718976395775952256 &  09:03:27.09 &  $ + $37:50:27.52 &   2.84 \\
                646255212109355136 &  09:32:43.76 &  $ + $26:59:18.71 &   5.43 \\
                799093375686189568 &  09:36:04.28 &  $ + $37:33:10.36 &    ... \\
                804739952015901056 &  10:22:10.56 &  $ + $41:13:46.31 &    ... \\
                3855208897392951808 &  10:28:55.55 &  $ + $00:50:27.58 &    ... \\
                5455707157211785216 &  10:43:28.27 &  $ - $29:03:51.43 &   6.90 \\
                3789271459953459200 &  11:04:41.47 &  $ - $04:13:15.91 &   6.06 \\
                761919883981626752 &  11:12:32.35 &  $ + $35:48:50.69 &   7.40 \\
                3585636855608873984 &  11:47:03.84 &  $ - $11:49:26.63 &    ... \\
                3901957795343984640 &  12:25:58.58 &  $ + $08:03:44.03 &    ... \\
                3520585968137789440 &  12:29:50.91 &  $ - $16:31:14.99 &    ... \\
                4010554050558613504 &  12:32:27.43 &  $ + $28:05:04.62 &    ... \\
                1717627794711189760 &  12:37:19.23 &  $ + $79:12:55.71 &   7.35 \\
                3957887649746831360 &  12:48:47.05 &  $ + $24:50:24.81 &    ... \\
                1472718211053416448 &  13:19:40.13 &  $ + $33:20:47.52 &    ... \\
                1565867461768654336 &  13:25:45.53 &  $ + $56:58:13.77 &   7.99 \\
                1484502295643717888 &  14:21:08.86 &  $ + $37:24:03.69 &   7.50 \\
                1668690628102524928 &  14:39:00.22 &  $ + $64:17:29.84 &   2.62 \\
                5882581895219920896 &  15:38:57.55 &  $ - $57:42:27.27 &   4.28 \\
                5989102478619520000 &  15:41:11.38 &  $ - $44:39:40.34 &   0.89 \\
                1222932018450870272 &  15:49:35.65 &  $ + $26:04:06.21 &    ... \\
                4503423641091792896 &  17:55:44.89 &  $ + $18:30:01.37 &    ... \\
                6721432232656219136 &  18:12:21.39 &  $ - $43:26:41.43 &  13.50 \\
                2099385000742174720 &  19:16:22.09 &  $ + $38:08:01.43 &    ... \\
                6643589352010758144 &  19:22:58.94 &  $ - $54:32:16.98 &   1.52 \\
                6794047652729200640 &  20:45:09.53 &  $ - $31:20:27.24 &   4.83 \\
                6456232811155783680 &  20:57:22.44 &  $ - $59:04:33.46 &  10.08 \\
                6409848126430575616 &  21:44:30.12 &  $ - $60:58:38.88 &   4.49 \\
                2224502448957645312 &  21:45:52.64 &  $ + $70:20:53.03 &   8.12 \\
                6564091190988411904 &  21:48:15.75 &  $ - $47:18:13.01 &    ... \\
                6410766630956403712 &  21:50:23.79 &  $ - $58:18:18.18 &    ... \\
                6408937971321777152 &  21:52:09.72 &  $ - $62:03:08.50 &   0.96 \\
                1898230241798771968 &  21:54:45.04 &  $ + $32:19:42.86 &    ... \\
                1999733272633122304 &  22:20:07.03 &  $ + $49:30:11.76 &   2.53 \\
                6628926642897745920 &  22:34:41.64 &  $ - $20:42:29.56 &    ... \\
                2207738916728293888 &  23:06:04.85 &  $ + $63:55:34.41 &   2.83 \\
                2282846074981010688 &  23:19:26.64 &  $ + $79:00:12.67 &   2.72 \\
                2395031273585836032 &  23:32:49.40 &  $ - $16:50:44.30 &    ... \\
                6387058411482256384 &  23:39:39.49 &  $ - $69:11:44.88 &   3.56 \\
                \hline
        \end{longtable}
        \tablefoot{
                The associated errors for the rotational periods are around 10\% of their value.
        }

\end{document}